\documentclass[nofootinbib,twocolumn,showpacs,preprintnumbers,pre,aps]{revtex4-1}

\usepackage[dvipdfmx]{graphicx}% Include figure files eps, pdf etc.
\usepackage{bm}% bold math
\usepackage{amsmath}
\usepackage{amssymb}
\usepackage{color}
\begin{document}

\title{Most probable path of an active Brownian particle}%

\author{Kento Yasuda}\email{yasudak@kurims.kyoto-u.ac.jp}
\author{Kenta Ishimoto}\email{ishimoto@kurims.kyoto-u.ac.jp}
\affiliation{
Research Institute for Mathematical Sciences, Kyoto University, Kyoto 606-8502, Japan}

\date{\today}

\begin{abstract}
In this study, we investigate the transition path of a free active Brownian particle (ABP) on a two-dimensional plane between two given states.
The extremum conditions for the most probable path connecting the two states are derived using the Onsager--Machlup integral and its variational principle.
We provide explicit solutions to these extremum conditions and demonstrate their nonuniqueness through an analogy with the pendulum equation indicating possible multiple paths.
The pendulum analogy is also employed to characterize the shape of the globally most probable path obtained by explicitly calculating the path probability for multiple solutions. 
We comprehensively examine a translation process of an ABP to the front as a prototypical example. Interestingly, the numerical and theoretical analyses reveal that the shape of the most probable path changes from an I to a U shape and to the $\ell$ shape with an increase in the transition process time.
The Langevin simulation also confirms this shape transition.
We also discuss further method applications for evaluating a transition path in rare events in active matter.
\end{abstract}
%\pacs{}

%\keywords{Suggested keywords}%Use show keys class option if keyword
\maketitle

\section{Introduction}
\label{Intro}

Active matter, such as a flock of birds, a school of fish, and bacteria, has attracted significant study interest in statistical mechanics in the last 20 years~\cite{Gompper20,LaugaBook, Shaebani20}.
Accordingly, several active particle models have been proposed to numerically reproduce a collective behavior.
One of the simplest models is the active Brownian particle (ABP) modeled using the Langevin equations. In an ABP, a particle moves at a constant speed along a randomly changing direction~\cite{Romanczuk12}.
Interestingly, active matter collective behaviors, such as motility-induced phase separation, have been examined using ABP models~\cite{Fily12,Cates15}.
The statistical properties of a single ABP, including self-diffusion and the hydrodynamic interactions between particles and a wall, have also been investigated~\cite{Patch17,Schaar15}.

Among the various active particles, biological and artificial microswimmers, such as bacteria and self-propelled Janus particles, have been intensively studied~\cite{Guasto12, Elgeti15, Goldstein15, Lauga16, Bees20, Gaffney21, Iwasawa21}.
As indicated by the scallop theorem~\cite{Purcell77,Shapere89}, the surrounding fluid of these microswimmers limits their motility and results in an abundantly dynamic behavior that often necessitates the reproduction of precise numerical calculations~\cite{Ishimoto13,Ishimoto17,Ohmura18,Ito19}.
The hydrodynamic effects are often masked under strong fluctuations by noisy environments, such as the simple thermal fluctuations generated by the fluctuation--dissipation theorem~\cite{KuboBook,DoiBook}. 
In addition, active system fluctuations are intrinsic and essential, as observed in bacterial run-and-tumble motions~\cite{LaugaBook} and the noisy background flow field induced by the surrounding active particles.

Herein, we focus on the transition of a stochastic active particle from the initial position, $\mathbf x_\mathrm i$, to an arbitrary final position, $\mathbf x_\mathrm f$, with time, $t_\mathrm f$, and consider the conditional probability, $\mathcal P(\mathbf x_\mathrm f,t_\mathrm f|\mathbf x_\mathrm i)$, of the transition.
The $\mathbf x_\mathrm i\to\mathbf x_\mathrm f$ transition may be a rare event when the conditional probability is minimal.
Although these rare transitional events have minor probabilities, they are essential for the survival of microswimmers because such events may result in the diversification of their habitats.
One of the major problems associated with these rare events is the extraction of the most probable path~\cite{Durr,Wissel79,Adib08,Wang10,Gladrow,Faccioli06,Yasuda22}. This path is the transition path exhibiting the highest path probability among the paths connecting the given initial and final states.
Alternatively, the most probable path is a \textit{typical} path of an \textit{atypical} transition with a tiny conditional probability $\mathcal P(\mathbf x_\mathrm f,t_\mathrm f|\mathbf x_\mathrm i)$.

Theoretical concepts, such as the path probability and the Onsager--Machlup (OM) integral, can be used to calculate the most probable path for the arbitrary initial and final states~\cite{Onsager53,RiskenBook,ZuckermanBook,Doi19}.
The most probable path of the transitions in case of a simple double-well potential is investigated through numerical calculations~\cite{Adib08} and experimental observations~\cite{Gladrow}. Several researchers have discussed the structural transitions of protein folding~\cite{ZuckermanBook,Faccioli06,Yasuda22}.
Notably, a chemical kinetic model was analyzed using the most probable path~\cite{Wang10}.

Further, several studies have introduced the OM integral for an active matter system~\cite{Wang21,Cates21,Majumdar20,Woillez19,Nardini17,Gu20} and used it to calculate the conditional probability of the ABP at short times~\cite{Majumdar20} and the escape rate of the run-and-tumble particle under trapping potential~\cite{Woillez19} with saddle-point approximation.
In addition, the most probable path of the ABP embedded in case of double-well potential is numerically calculated under a small translational noise limit~\cite{Gu20}.
Nonetheless, the abovementioned studies could not fully understand the exact shape of the transition path because the OM integral minimizer obeys nonlinear equations, with solutions not as unique as those discussed herein do.
Our study provides analytical solutions, calculates the most probable path, and classifies its shape for the transition process of a free ABP.

Here, we derive the path probability written by the OM integral from the Langevin equations for the positions and orientation of the ABP.
We deduce the most probable path using the variational principle of the OM integral because the minimum OM integral results in the maximum path probability.
We also discover that these extremum conditions for the most probable path are analogous to the pendulum equation, enabling formal analytical solutions.
However, analytically determining the unknown coefficients for arbitrary boundary conditions are not feasible; hence, we numerically resolve the equations to obtain multiple solutions from a unique boundary condition.
Finally, we demonstrate the exact shape of the most probable path 
in the case of a translation to the front.

We also review the original equations of the single ABP in the next section.
Subsequently, the OM integral and the derivation of the extremum conditions for the most probable path are explained in Section~\ref{OMVPsec}.
Their analytical solutions are presented in Section~\ref{Asol}. The numerical solutions for the specific boundary condition are discussed in Section~\ref{MPPsol}.
Section~\ref{Sec:PhaseDiagram} describes the phase diagram of the most probable path shape for various boundary condition values.
Finally, Section~\ref{Dis} provides the summary and discussions.

\section{Active Brownian particle}
\label{Model}

\begin{figure}[t]
\begin{center}
\includegraphics[scale=0.35]{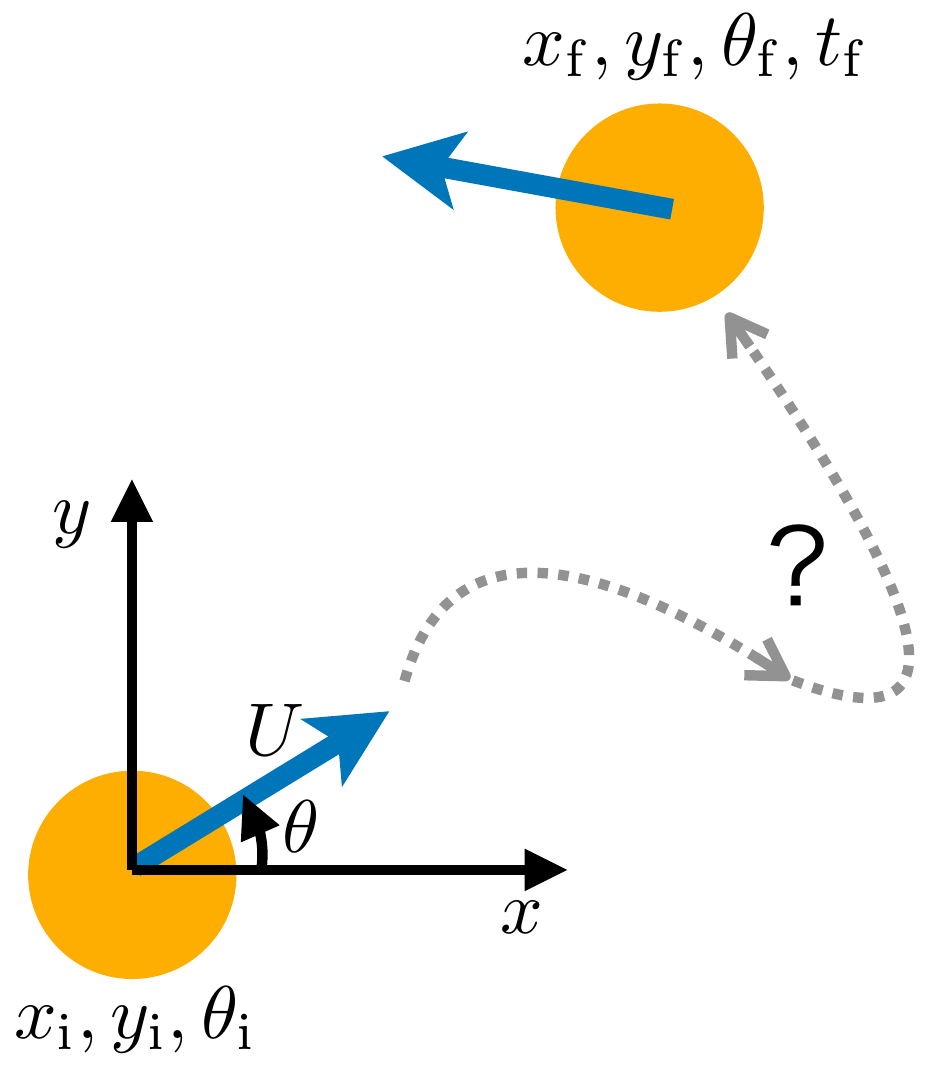}
\end{center}
\caption{
An ABP is at $(x,y)$ in the 2D space where it moves with a constant propulsion speed, $U$, and an orientation, $\theta$, which is the angle from the $x$ axis. The position and the orientation randomly fluctuate under thermal or active noise.
Therefore, the single ABP dynamics is written as shown in Eqs.~(\ref{LE-x})--(\ref{LE-th}).
The rare events, where the ABP is obtained at $x_\mathrm i,y_\mathrm i,\theta_\mathrm i$ at the initial time and $x_\mathrm f, y_\mathrm f, \theta_\mathrm f$ at the final time $t_\mathrm f$, are analyzed using the OM integral given in Eq.~(\ref{OMint}) and the most probable path obeying Eqs.~(\ref{MPPE-x})--(\ref{MPPE-th}).}
\label{Fig:model}
\end{figure}

This section summarizes the model equation for an ABP, which is a simple, but canonical model for an active particle.
Let us consider an ABP moving in a two-dimensional (2D) space $(x,y)$, as shown in Fig.~\ref{Fig:model}.
The ABP actively moves along the particle orientation, $\theta$, with a constant propulsion speed, $U$.
The orientation, which is the $\theta$ (and position $x$ and $y$), experiences the noise caused by its activity or the thermal motions of a surrounding fluid.
Therefore, this orientation noise results in the angle diffusion of the ABP.
The ABP position $x(t)$, $y(t)$ and orientation $\theta(t)$ dynamics are described using the following Langevin equations \cite{Shaebani20}:
\begin{align}
&\dot x=U\cos \theta+\xi_x(t),
\label{LE-x}\\
&\dot y =U\sin \theta+\xi_y(t),
\label{LE-y}\\
&\dot \theta=\xi_\theta(t).
\label{LE-th}
\end{align}
where the dot represents the time derivative.
$\xi_\alpha(t)$ ($\alpha, \beta=x, y, \theta$) is the Gaussian white noise satisfying conditions $\langle\xi_\alpha\rangle=0$ and $\langle\xi_\alpha(t)\xi_\beta(0)\rangle =2D_{\alpha\beta}\delta(t)$, where $D_{\alpha\beta}$ is a positive definite diffusion tensor.

Here, we assumed that an isotropic diffusion tensor, that is, $D_{xx}=D_{yy}=D_\mathrm t$, $D_{\theta\theta}=D_\mathrm r$, and the other components vanish.
In the case of the thermal noise for a sphere with a radius, $a$, in a viscous fluid with viscosity, $\eta$, we deduced $D_\mathrm t=k_\mathrm BT/(6\pi\eta a)$ and $D_\mathrm r=k_\mathrm BT/(8\pi\eta a^3)$~\cite{DoiBook,KKbook}, where $k_\mathrm B$ is the Boltzmann constant, and $T$ is a temperature characterizing the noise magnitude.
In the Appendix~\ref{App:ellip}, an anisotropic diffusion tensor (e.g., $D_{xx}\ne D_{yy}$) is considered for an ellipsoid-shaped ABP.

Eqs.~(\ref{LE-x})--(\ref{LE-th}) represent a single free ABP.
Therefore, we generalized it to more complicated cases, where the potential force and the background flow are considered by replacing the drift velocity.
Section~\ref{Dis} discusses this generalized case.

\section{Onsager--Machlup variational principle}
\label{OMVPsec}
Let us consider the following situation: we determined the ABP at the initial state $x_\mathrm i, y_\mathrm i, \theta_\mathrm i$ at $t=0$ and the final state $x_\mathrm f, y_\mathrm f, \theta_\mathrm f$ at $t=t_\mathrm f$.
This transition from the initial to the final state is a rare event emerging from the noise. The problem considered here is the most probable transition path between these two states [Fig.~\ref{Fig:model}].

This section presents a framework for calculating the most probable path of the ABP using the OM integral and its variational principle.
This principle leads to equations determining the most probable path for the transition between the arbitrary initial and final states.

\subsection{Onsager--Machlup integral}

The path probability, $P[x(t),y(t),\theta(t)]$, which is the probability of a specific stochastic trajectory, is used to analyze stochastic systems~\cite{RiskenBook}.
Let us set the initial condition as $x(0)=x_\mathrm i$, $y(0)=y_\mathrm i$ and $\theta(0)=\theta_\mathrm i$. The path probability $P[x(t),y(t),\theta(t)|x_\mathrm i, y_\mathrm i, \theta_\mathrm i]$ during the time interval $0\le t<t_\mathrm f$ is then given by $P[x,y,\theta|x_\mathrm i, y_\mathrm i, \theta_\mathrm i]=C\exp(-O[x,y,\theta]/2k_\mathrm BT)$~\cite{Onsager53,RiskenBook}.
Here, $C$ is a normalization constant determined by condition $\int\mathcal D \mathbf x\,P[x,y,\theta|x_\mathrm i, y_\mathrm i, \theta_\mathrm i]=1$, where $\int\mathcal D \mathbf x$ indicates integration over all paths for $x(t), y(t), \theta(t)$,
and $O[x(t),y(t),\theta(t)]$ is the OM integral derived as follows: 
\begin{align}
&O[x(t),y(t),\theta(t)]=\frac{k_\mathrm BT\,\mathrm{Pe}}{2}\int_{0}^{t_\mathrm f}\frac{dt}{\tau}\,\Bigl[(\dot{\bar x}\tau - \cos\theta)^2\nonumber\\
&+(\dot{\bar y}\tau - \sin\theta)^2+\dot \theta^2\tau^2\Bigr]
\label{OMint}
\end{align}
using Eqs.~(\ref{LE-x})--(\ref{LE-th}). The detailed derivations of the above equations are provided in the Appendix~\ref{Apps:PathProb}.
Note that the OM integral formulation possesses an indeterminacy issue caused by various possible forms of time discretization. However, this indeterminacy does not affect the case of the single free ABP (Appendix~\ref{Apps:PathProb}).
The abovementioned equations contained a nondimensional position, $\bar x=x/L,\bar y=y/L$, where $L=\sqrt{D_\mathrm t/D_\mathrm r}$ is the length scale representing a particle size.
Here, the rotational P\'eclet number, $\mathrm{Pe}=(D_\mathrm r\tau)^{-1}$~\cite{Patch17}, representing the noise to mobility ratio and the time scale $\tau=L/U$, which the particle spends traveling its body size, are also introduced. 
In a real bacterial system, the length scale, $L$, and the time scale, $\tau$, are estimated as $L\sim 10\mu \mathrm m$ and $\tau\sim 1\mathrm s$, respectively~\cite{LaugaBook,Purcell77}.

\subsection{Onsager--Machlup variational principle}
The OM variational principle states that the transition path minimizing the OM integral has the highest probability. 
Conversely, the most probable path can be obtained by requiring a positive disappearance of the first and second variations of the OM integral (i.e., $\delta O=0$ and $\delta^2O>0$, respectively). Considering the first variation of Eq.~(\ref{OMint}) concerning $x(t)$, $y(t)$, and $\theta(t)$ yields the following extremum conditions for the most probable path:
\begin{align}
&\ddot{\bar x}=-\tau^{-1}\dot \theta\sin\theta,
\label{MPPE-x}\\
&\ddot{\bar y}= \tau^{-1}\dot \theta\cos\theta,
\label{MPPE-y}\\
&\ddot\theta=\tau^{-1}(\dot{\bar x}\sin\theta-\dot{\bar y}\cos\theta).
\label{MPPE-th}
\end{align}
The detailed derivations of the abovementioned equations are obtained in the Appendix~\ref{App:DerEq}.
The positive second variation, $\delta^2O>0$, is also required for the minimum OM integral, $O[\mathbf x(t)]$.
We can confirm that the Legendre conditions~\cite{CourantBook}, which are a necessary condition for $\delta^2O>0$, always hold with a positive definite diffusion matrix, $D_{\alpha\beta}$, as assumed in the previous section.
Specifically, $\delta O=0$ and $\delta^2O>0$ are the only conditions for the \textit{local} minimum path.
Hence, we will further compare the OM integral of each solution to the extremum conditions to determine the \textit{global} minimum path from multiple local minimum paths.

Two boundary conditions were required to solve the second-order differential equations, Eqs.~(\ref{MPPE-x})--(\ref{MPPE-th}).
We employed the Dirichlet (or first type) boundary condition represented by the initial condition $x_\mathrm i, y_\mathrm i, \theta_\mathrm i$ at $t=0$ and the final condition $x_\mathrm f, y_\mathrm f, \theta_\mathrm f$ at $t=t_\mathrm f$.
We set the initial condition as $x_\mathrm i=0, y_\mathrm i=0, \theta_\mathrm i=0$ without loss of generality because the system has translational and rotational invariance. 
The parameters in this problem are only the final conditions, $x_\mathrm f, y_\mathrm f, \theta_\mathrm f$, and the final time, $t_\mathrm f$, providing the system time and length scales.

Recall that the path probability is given by an exponential of the OM integral as $P\sim \exp(-\mathrm{Pe}\,\hat O/4)$, where we used $\hat O=2O/(k_\mathrm BT\,\mathrm{Pe})$. 
The path probability in the small noise limit, $\mathrm{Pe}\to\infty$, converges to the most probable path, $\mathbf x^\mathrm{MPP}$. The probabilities for the other transition paths then become zero~\cite{EllisBook}. 
In this limit, the path-averaged value of a functional $A[\mathbf x(t)]$ may be approximated by that of the most probable path as $\langle A\rangle_{\mathrm i\to\mathrm f}\approx A[\mathbf x^\mathrm{MPP}(t)]$, when the $\mathrm{Pe}$ dependence on $A$ is weaker than the exponential function (i.e., $\ln A= o(\mathrm{Pe})$).

\subsection{Entropy change}
We evaluated the entropy change of the thermal bath along the trajectory, $\Delta s_\mathrm b[x(t),y(t),\theta(t)]$, as follows according to the fluctuation theorem~\cite{Seifert12}:
\begin{align}
\frac{P[x(t),y(t),\theta(t)|x_\mathrm i, y_\mathrm i, \theta_\mathrm i]}{P[x^{\mathrm {rev}}(t),y^{\mathrm {rev}}(t),\theta^{\mathrm {rev}}(t)|x_\mathrm f, y_\mathrm f, \theta_\mathrm f]}=e^{\Delta s_\mathrm b/k_\mathrm B},
\label{eq:FT}
\end{align}
where $x^{\mathrm {rev}}$ is the reversed path defined as $x^{\mathrm {rev}}(t)=x(t_\mathrm f-t)$.
Substituting Eq.~(\ref{OMint}) to Eq.~(\ref{eq:FT}), we derive an explicit form of $\Delta s_\mathrm b$, which is given as follows:
\begin{align}
\Delta s_\mathrm b[x(t),y(t),\theta(t)]=k_\mathrm B\,\mathrm{Pe}\int_{0}^{t_\mathrm f}dt\,\left[\dot{\bar x}\cos\theta+\dot{\bar y}\sin\theta\right].
\label{EP}
\end{align}
We evaluated the entropy change of the thermal bath or the irreversibility of the most probable path using the abovementioned derived formula.

\section{Analytical treatment with pendulum analogy}
\label{Asol}

\begin{figure}[t]
\begin{center}
\includegraphics[scale=0.6]{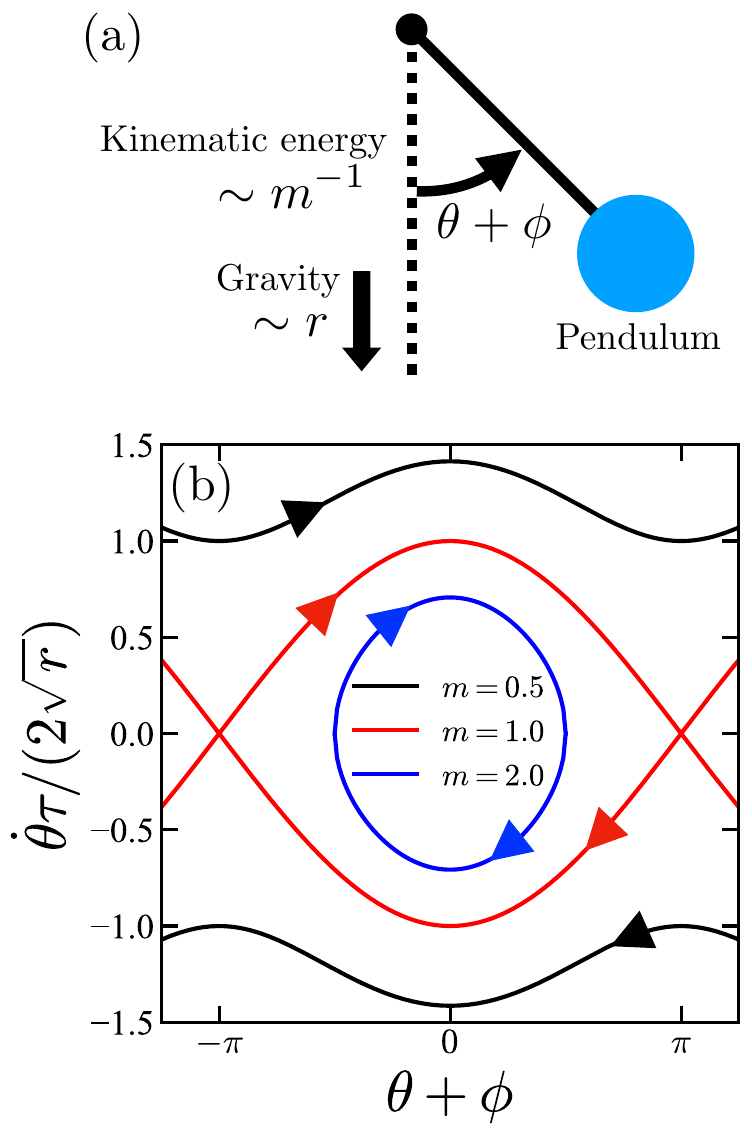}
\end{center}
\caption{(a) Schematic of the pendulum system corresponding to the extremum condition for $\theta$, Eq.~(\ref{eq-pen}). 
The angle from an equilibrium state is given by $\theta(t)+\phi$.
A nondimensional parameter, $r$, characterizes the downward gravity magnitude.
The kinematic energy, which is a conservative quantity, is characterized by a nondimensional parameter, $m$.
Three parameters, namely $\phi$, $r$, and $m$, are determined using three final conditions, that is, $x_\mathrm f$, $y_\mathrm f$, and $\theta_\mathrm f$. 
(b) Phase diagram of the $\theta$ calculated from Eq.~(\ref{sol-th}) showing the renowned pendulum dynamics.
Different colors indicate various $m$ values.
Dynamics constantly evolve in the clockwise direction, as shown by the arrows.
In the case of $m=0.5$ (black line), the trajectory exhibits ``rotation" dynamics, which can be observed when $m<1$.
Therefore, the critical behavior emerges with $m = 1$ (red line), including fixed points (i.e., $\dot\theta=0$ and $\theta+\phi=(2n+1)\pi$).
Conversely, where $m=2.0$ (blue line), the trajectory exhibits ``swing" dynamics that may include periodic cycles characterized by the period, $T_\mathrm c$. However, the ``swing" dynamics is only possible when $m > 1$.
}
\label{Fig:PS}
\end{figure}

The most probable path can be obtained by solving the extremum conditions (Eqs.~(\ref{MPPE-x})--(\ref{MPPE-th})) with boundary conditions.
Although these equations are nonlinear differential, they can be formally reduced to the pendulum's equation of motion. A general solution may then be deduced in an analytical form.
This section discusses the most probable path using these analytical treatments.

\subsection{Equations of $x$, $y$, and $\theta$}

First, we considered Eqs.~(\ref{MPPE-x}) and (\ref{MPPE-y}) for $\bar x$ and $\bar y$, respectively. These equations are formally solved as follows:
\begin{align}
&\bar x=\int_0^t\frac{dt'}{\tau}\,\cos\theta(t')+(\bar V_x-1)\frac{t}{\tau},
\label{sol-x}\\
&\bar y =\int_0^t\frac{dt'}{\tau}\,\sin\theta(t')+\bar V_y\frac{t}{\tau}.
\label{sol-y}
\end{align}
We used the initial conditions $x_\mathrm i=0$ and $y_\mathrm i=0$ at $t=0$.
$\bar V_x$ and $\bar V_y$ represent the nondimensional initial velocities $\dot{\bar x}(0)\tau$ and $\dot {\bar y}(0)\tau$, respectively, and must be decided by the final conditions $\bar x_\mathrm f=x_\mathrm f/L$ and $\bar y_\mathrm f=y_\mathrm f/L$, respectively, as follows:
\begin{align}
&\bar V_x=1+\frac{\bar x_\mathrm f \tau}{t_\mathrm f}-\frac{\tau}{t_\mathrm f}\int_0^{t_\mathrm f}\frac{dt'}{\tau}\,\cos\theta(t'),
\label{Vx}\\
&\bar V_y =\frac{\bar y_\mathrm f\tau}{t_\mathrm f}-\frac{\tau}{t_\mathrm f}\int_0^{t_\mathrm f}\frac{dt'}{\tau}\,\sin\theta(t').
\label{Vy}
\end{align}
These expressions show that $\bar V_x$ and $\bar V_y$ depend on the dynamics of $\theta$ in the entire time from $t=0$ to $t_\mathrm f$.

Using Eqs.~(\ref{sol-x}) and (\ref{sol-y}), the dynamics of $\theta$, (Eq.~(\ref{MPPE-th})) is rewritten as follows:
\begin{align}
&\ddot\theta\tau^2=(\bar V_x-1)\sin\theta-\bar V_y\cos\theta=-r\sin(\theta+\phi),
\label{eq-pen}
\end{align}
where we used $r=\sqrt{(\bar V_x-1)^2+\bar V_y^2}$, $\cos\phi=-(\bar V_x-1)/r$, and $\sin\phi=\bar V_y/r$.
This equation is entirely similar to the renowned pendulum equation~\cite{Belendez07}, where $r$ and $\theta(t)+\phi$ correspond to the gravity force magnitude and the pendulum angle, respectively [Fig.~\ref{Fig:PS}(a)].
When $\phi$ is an integer multiple of $\pi$, $\theta=0$ becomes a trivial solution to this equation.
When $\phi$ is given by an even multiple of $\pi$, this trivial solution becomes stable.
However, this solution becomes unstable when $\phi$ is an odd multiple of $\pi$.

Therefore, multiplying Eq.~(\ref{eq-pen}) by $\dot\theta$ and integrating once deduce the following nontrivial solution:
\begin{align}
&\dot\theta\tau=\pm2\sqrt{\frac{r}{m}}\sqrt{1-m\sin^2((\theta+\phi)/2)},
\label{sol-th}
\end{align}
where $m$ is a positive coefficient determined by the boundary conditions, which denotes the inverse of the pendulum's kinematic energy.
For $0<m<1$, $\theta(t)$ monotonically increases or decreases with time. This behavior is known as ``rotation."
Alternatively, for $1<m$, $\theta(t)$ oscillates with its period for one cycle, $T_\mathrm c$.
This behavior is known as a ``swing," which indicates an analog to the pendulum dynamics.
In the ``swing" dynamics, $\theta$ is bound as $|\theta+\phi|<\Theta_{\mathrm {max}}$ with a finite amplitude:
\begin{align}
\Theta_{\mathrm {max}}=2\sin^{-1}(m^{-1/2}).
\label{Eq:thmax}
\end{align}
The solution to Eq.~(\ref{sol-th}) is plotted in the phase space in Fig.~\ref{Fig:PS}(b).

\subsection{Passage time}

\begin{figure}[!t]
\begin{center}
\includegraphics[scale=0.5]{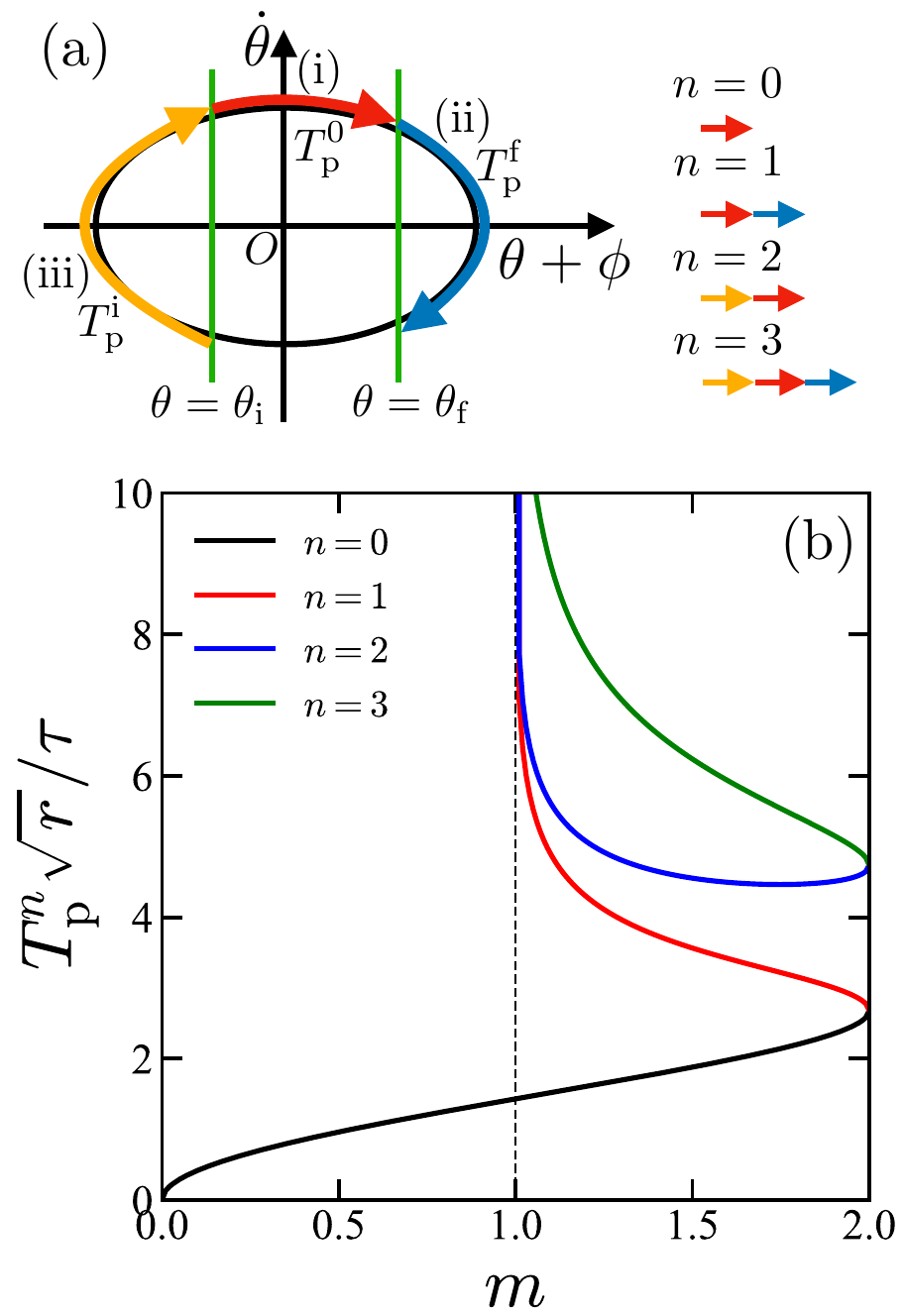}
\end{center}
\caption{(a) Various passage processes from $\theta=\theta_\mathrm i=0$ to $\theta=\theta_\mathrm f$ shown in the phase space spanned by $\theta+\phi$ and $\dot \theta$. 
The green vertical lines indicate the initial ($\theta=\theta_\mathrm i=0$) and final ($\theta=\theta_\mathrm f$) states.
The process (i) is the shortest process for $\theta_\mathrm i\to\theta_\mathrm f$ (indicated by the red arrow). The duration is denoted by $T_\mathrm p^0$ in Eq.~(\ref{eq-Tp0}).
Conversely, the recurrent process (ii) is for $\theta_\mathrm f\to\theta_\mathrm f$ (indicated by the blue arrow), with its duration, $T_\mathrm p^\mathrm f$, being presented in Eq.~(\ref{eq-Tpf}). 
The yellow arrow indicates the recurrent process (iii) for $\theta_\mathrm i\to\theta_\mathrm i$. The passage time is $T_\mathrm p^\mathrm i$ in Eq.~(\ref{eq-Tpi}).
Note that $T_\mathrm p^\mathrm f<T_\mathrm p^\mathrm i$ in this figure.
The $n=0$ process for $\theta_\mathrm i\to\theta_\mathrm f$ corresponds to the curve indicated by the red arrow; the $n=1$ process is represented by the red and blue arrows; the $n=2$ process is depicted by the yellow and red arrows; and finally, the $n=3$ process is represented by the sum of the three arrows.
(b) Passage time, $T_\mathrm p^n(m)$, computed from Eqs.~(\ref{Tp4ell})--(\ref{Tp4ellp3}) as a function of $m$ (i.e., approximately $n=3$). 
We set $\phi=-\pi/3$ and $\theta_\mathrm f+\phi=\pi/2$; therefore, $m_\mathrm{max}=2$, as indicated by Eq.~(\ref{mmax}).
$T_\mathrm p^n(m)$ depends on the passage process labeled as $n$ ((a)).
A different color plot shows each passage time, $T_\mathrm p^n(m)$, for different processes.
The black vertical dashed line represents $m=1$, which is the critical value from the ``rotation" to the ``swing" dynamics.
Only $T_\mathrm p^0$ (black curve) has a finite value across the range $0<m<1$ because the remaining passage times diverge at $m=1$.
}
\label{Fig:Tp}
\end{figure}

The passage time, $T_\mathrm p(m)$, which is a characteristic time associated with the ``swing" dynamics and obtained as a solution to Eq.~(\ref{sol-th}), is discussed below.
The passage time is defined as the duration from $\theta_\mathrm i=0$ to $\theta_\mathrm f$.
The process $\theta_\mathrm i\to\theta_\mathrm f$ cannot be uniquely determined because the pendulum can oscillate multiple times before reaching the final angle.
Let us consider the process mapped in the phase space to distinguish each passage process (Fig.\ref{Fig:Tp}(a)).
We constructed the passage process for an arbitrary choice of $\theta_\mathrm f$ and $\phi$ by categorizing the process $\theta_\mathrm i\to\theta_\mathrm f$ into four parts: (i) shortest process for $\theta_\mathrm i\to\theta_\mathrm f$ (red arrow in Fig.\ref{Fig:Tp}(a)); (ii) recurrent process for $\theta_\mathrm f\to\theta_\mathrm f$ (blue arrow in Fig.\ref{Fig:Tp}(a)); (iii) recurrent process for $\theta_\mathrm i\to\theta_\mathrm i$ (yellow arrow in Fig.\ref{Fig:Tp}(a)); and (iv) process that encloses one cycle. 
Subsequently, we introduced the partial passage time in each process. The following representations are easily obtained from the pendulum equation properties\cite{Belendez07}.
We spent the following passage time for (i): 
\begin{align}
T_\mathrm p^0(m)=\mathrm{Sign}(\theta_\mathrm f)\tau\sqrt{\frac{m}{r}}[F((\theta_\mathrm f+\phi)/2,m)-F(\phi/2,m)],
\label{eq-Tp0}
\end{align}
where we used a sign function, $\mathrm{Sign}(z)=1~(z\ge0), -1~(z<0)$ [notice $\mathrm{Sign}(0)=1$], and the incomplete elliptic integral of the first kind:
\begin{align}
F(\psi,k)=\int_0^\psi\frac{dz}{\sqrt{1-k\sin^2z}}.
\end{align}
The passage times for recurrent processes (ii) and (iii) are respectively given as follows:
\begin{align}
&T_\mathrm p^\mathrm f(m)=T_\mathrm c(m)/2-2\,\mathrm{Sign}(\theta_\mathrm f)\tau\sqrt{\frac{m}{r}}F((\theta_\mathrm f+\phi)/2,m),
\label{eq-Tpf}
\end{align}
and 
\begin{align}
&T_\mathrm p^\mathrm i(m)=T_\mathrm c(m)/2+2\,\mathrm{Sign}(\theta_\mathrm f)\tau\sqrt{\frac{m}{r}}F(\phi/2,m),
\label{eq-Tpi}
\end{align}
where $T_\mathrm c(m)$ is the time for one cycle of the swinging pendulum that characterizes process (iv) given as follows:
\begin{align}
T_\mathrm c(m)&=4\tau\sqrt{\frac{1}{r}}F(\pi/2,1/m).
\label{eq-Tc}
\end{align}
Note that $T_\mathrm c(m)$ has a lower bound as $T_\mathrm c(m)> 2\pi\tau/\sqrt{r}$ and diverges as $T_\mathrm c(1)\to\infty$.

We next construct multiple passage times for $\theta_\mathrm i\to\theta_\mathrm f$ by combining the four partial passage times of $T_\mathrm p^0$, $T_\mathrm p^\mathrm i$, $T_\mathrm p^\mathrm f$, and $T_\mathrm c$.
Each passage time is labeled in order from the smallest and defined as $T_\mathrm p^n$ ($n=0,1,2,\cdots$).
The $n=0$ passage time, $T_\mathrm p^0$, is the shortest process, as defined in Eq.~(\ref{eq-Tp0}) and indicated by the red arrow in Fig.~\ref{Fig:Tp}(a).
We first consider the case when $T_\mathrm p^\mathrm f<T_\mathrm p^\mathrm i$ (Fig.~\ref{Fig:Tp}(a)). Using the definition of $T_\mathrm p^n$, the $n=1$ passage time, $T_\mathrm p^1$, is spent by the process constructed with processes (i) and (ii), which represents the sum of the red and blue arrows in Fig.~\ref{Fig:Tp}(a). Meanwhile, the $n=2$ passage time, $T_\mathrm p^2$, is made by combining processes (i) and (iii) indicated by the yellow and red arrows in Fig.~\ref{Fig:Tp}(a), respectively.
In the reversed order of the two recurrent passage times (i.e., $T_\mathrm p^\mathrm f>T_\mathrm p^\mathrm i$), the $T_\mathrm p^1$ and $T_\mathrm p^2$ processes are exchanged to generate the order from the smallest following the definition. 
Irrespective of the size of the recurrent passage times, the $n=3$ passage time, $T_\mathrm p^3$, includes the three processes of (i) to (iii) and is represented by the sum of the yellow, red, and blue arrows in Fig.~\ref{Fig:Tp}(a).
These processes labeled from $n=0$ to $n=3$ constitute the bases of higher-order processes because all passages are created by one of the four shortest passages and the additional cycles characterized by time, $T_\mathrm c(m)$.
For example, the $n=4$ process can be constructed using the $n=0$ process and an entire cycle (i.e., $T_\mathrm p^4=T_\mathrm p^0+T_\mathrm c$).
We constructed the $n$-th passage time $T_\mathrm p^n(m)$ as follows based on the abovementioned statements:
\begin{align}
&T_\mathrm p^{4\ell}(m)=T_\mathrm p^0(m)+\ell T_\mathrm c(m),
\label{Tp4ell}\\
&T_\mathrm p^{4\ell+1}(m)=T_\mathrm p^0(m)+\ell T_\mathrm c(m)+\mathrm{min}[T_\mathrm p^\mathrm i(m),T_\mathrm p^\mathrm f(m)],
\label{Tp4ellp1}\\
&T_\mathrm p^{4\ell+2}(m)=T_\mathrm p^0(m)+\ell T_\mathrm c(m)+\mathrm{max}[T_\mathrm p^\mathrm i(m),T_\mathrm p^\mathrm f(m)],
\label{Tp4ellp2}\\
&T_\mathrm p^{4\ell+3}(m)=T_\mathrm p^0(m)+\ell T_\mathrm c(m)+T_\mathrm p^\mathrm i(m)+T_\mathrm p^\mathrm f(m),
\label{Tp4ellp3}
\end{align}
where $\ell=0,1,2,\cdots$ denotes the number of cycles in the corresponding process.
The above expressions for the passage time are available, even in the case of $\theta_\mathrm f<0$ compared to Fig.~\ref{Fig:Tp}(a). 
We obtained $T_\mathrm p^0=0$ and $T_\mathrm p^\mathrm L(m)+T_\mathrm p^\mathrm R(m)=T_\mathrm c(m)$ when $\theta_\mathrm f=0$. Therefore, several passage times degenerated as $T_\mathrm p^{4\ell-1}=T_\mathrm p^{4\ell}$.
To realize the process $\theta_\mathrm i=0\to\theta_\mathrm f$, the parameter $m$ must satisfy $1<m<m_\mathrm {max}$, where
\begin{align}
m_\mathrm {max}=
\begin{cases}
1/\sin^2(\phi/2) &~(|\phi|\ge|\theta_\mathrm f+\phi|)\\
1/\sin^2((\theta_\mathrm f+\phi)/2) &~(|\phi|<|\theta_\mathrm f+\phi|)
\end{cases}.
\label{mmax}
\end{align}

Compared to the ``swing" dynamics, which enables multiple passage processes, the ``rotation" dynamics ($0<m<1$) only permits a single passage process with time, $T_\mathrm{p}^0(m)$.

Fig.~\ref{Fig:Tp}(b) plots the passage time, $T_\mathrm p^n(m)$, as a function of $m$ for a particular parameter set. This figure clearly shows that, for a given final condition,
\begin{align}
T_\mathrm p^n(m)=t_\mathrm f,
\end{align}
with $t_\mathrm f$ as a sufficiently large value, 
the multiple values of $m$ are possible solutions. Multiple solutions to the extremum conditions can specifically exist for a given boundary condition.

\section{Demonstrations of the most probable path}
\label{MPPsol}

\begin{figure*}[!t]
\begin{center}
\includegraphics[scale=0.4]{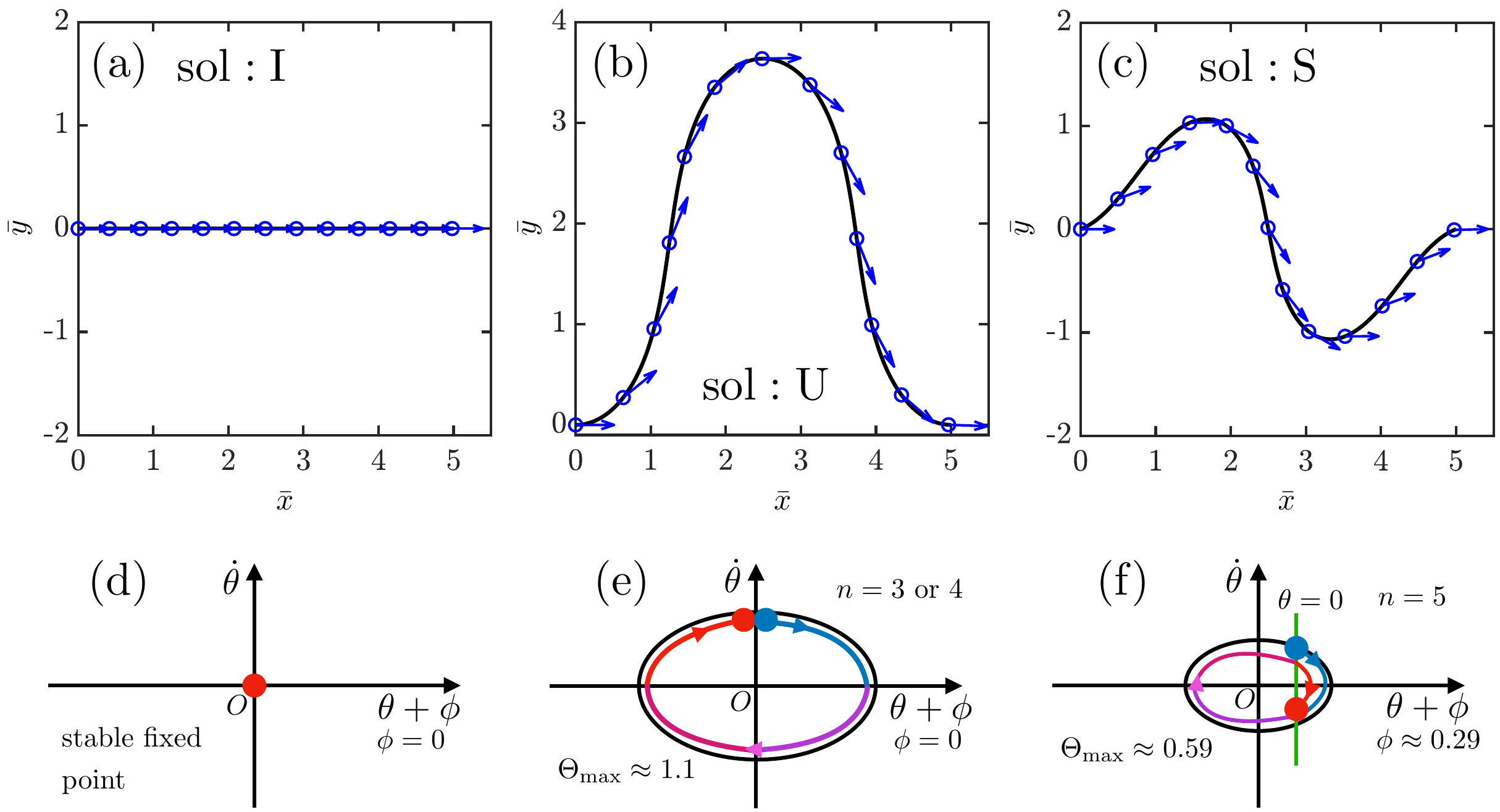}
\end{center}
\caption{Three independent solutions to the extremum conditions, Eqs.~(\ref{MPPE-x})--(\ref{MPPE-th}), that is, (a) sol:I, (b) sol:U, and (c) sol:S, plotted in the $\bar x$--$\bar y$ space under the same boundary conditions of $\bar x_\mathrm f=5,\bar y_\mathrm f=0,\theta_\mathrm f=0$, and $t_\mathrm f/\tau=12$.
The blue arrows indicate $\theta$ at each point. $\tau$ is the time interval of these arrows.
Parameters $r,\phi$, and $m$ are determined as follows for each solution: 
(a) $r\approx0.5833, \phi=0$ (or $\bar V_x\approx 0.4167, \bar V_y=0$), and $m\to\infty$,  
(b) $r\approx0.3115, \phi=0$ (or $\bar V_x\approx0.6885, \bar V_y=0$), and $m\approx3.7790$, 
(c) $r\approx0.5067, \phi\approx0.2857$ (or $\bar V_x\approx0.5138, \bar V_y\approx0.1428$), and $m\approx11.8167$.  
The OM integral and the entropy change of the thermal bath are estimated as follows:
(a) $\hat O=49/12\approx4.0833$ and $\Delta \hat s_\mathrm b=5$,
(b) $\hat O\approx3.0887$ and $\Delta \hat s_\mathrm b\approx9.2896$, 
(c) $\hat O\approx4.0029$ and $\Delta \hat s_\mathrm b\approx6.4975$,
where $\hat O=2O/(k_\mathrm BT\,\mathrm{Pe})$ and $\Delta \hat s_\mathrm b=\Delta s_\mathrm b/(k_\mathrm B\,\mathrm{Pe})$.
(d--f) Schematic showing the time evolution in each solution's phase space of $\theta$. The trajectories are shown as arrows from the initial (blue point) to the final (red point) states. 
$\Theta_\mathrm{max}$, which is the maximum value of $\theta+\phi$, for each solution, is calculated from Eq.~(\ref{Eq:thmax}).
(d) sol:I is at the stable fixed point indicated by the red point.
(e) sol:U corresponds to the passage process $n=3$ or $4$ (i.e., entire cycle). 
(f) In sol:S, a nonzero $\phi$ is indicated by the green vertical line. The initial and final states are shifted from the vertical axis of $\theta+\phi=0$. 
The solution corresponds to the passage process $n=5$ comprising process (ii) and an additional cycle [Fig.~\ref{Fig:Tp}(a)].
}
\label{Fig:sol}
\end{figure*}

As previously discussed, the overall solutions for the extremum conditions above are presented in Eqs.~(\ref{sol-x}), (\ref{sol-y}), and (\ref{sol-th}). Hence, parameters $r,\phi$, and $m$ must be decided by the boundary conditions.
However, analytically determining the parameters (i.e., $r,\phi$, and $m$) is difficult for the arbitrary boundary conditions because parameters $r$ and $\phi$ depend on the integral of $\theta(t)$ over time (Eqs.~(\ref{Vx}) and (\ref{Vy})), while $m$ depends on $r$ and $\phi$.
This section numerically solves the extremum conditions (Eqs.~(\ref{MPPE-x})--(\ref{MPPE-th}), under some specific boundary conditions).

\subsection{Translation to the front}
Consider the most probable path for the forward transition, where the final state occurs before the initial state. As a typical and physically natural situation, we set the boundary conditions as $\bar x_\mathrm f=5,\bar y_\mathrm f=0,\theta_\mathrm f=0, t_\mathrm f/\tau=12$.
The zero final angles, $\theta_\mathrm f=0$, indicating that the ABP is in the same direction as the initial time are determined at the final time. This simple case is a classic example because a nontrivial particle trajectory is selected as the most probable path among numerous solutions of the extremum equation both with the ``swing" and ``rotation" dynamics of the angle variable. 

Three independent solutions are obtained by numerically resolving Eqs.~(\ref{MPPE-x})--(\ref{MPPE-th}) using a MATLAB solver \textsf{bvp4c} (Figs.~\ref{Fig:sol}(a)--(c)).
We represent the straight I-shaped solution in Fig.\ref{Fig:sol}(a) sol:I, U-shaped solution in (b) sol:U, and S-shaped solution in (c) sol:S.
Parameters $r,\phi$, and $m$ of each solution can be estimated by applying Eqs.~(\ref{Vx}), (\ref{Vy}), and (\ref{sol-th}), respectively, to the numerical solutions.
The estimated parameter values are provided in the caption of Figs.~\ref{Fig:sol}(a)--(c).
The numerical value of the OM integral and the entropy production are estimated using Eqs.~(\ref{OMint}) and (\ref{EP}), respectively, and are available in the caption of Figs.~\ref{Fig:sol}(a)--(c).

Fig.~\ref{Fig:Tp}(a) shows that the numerical values of $r,\phi$, and $m$ for each solution can predict the $\theta$ dynamics in the phase space.
Fig.\ref{Fig:sol}(d)--(f) provide a schematic of the $\theta$ dynamics in the phase space for sol:I, sol:U, and sol:S, respectively.
In the case of sol:I (Fig.\ref{Fig:sol}(d)), the solution remains at the origin of the phase space, which is a stable fixed point that corresponds to the pendulum at the stationary state condition (i.e., $\theta+\phi=0$ in Fig.~\ref{Fig:PS}(a)).
In the case of sol:U (Fig.\ref{Fig:sol}(e)), the solution shows the $n=3$ or $n=4$ ``swing" dynamics and satisfies the final condition after a single cycle.
Accordingly, $\phi=0$; thus, the initial and final conditions are the origin of the horizontal axis, $\theta+\phi$, in the phase space.
This situation can be compared to a pendulum flung at the bottom with a finite velocity and returns after one swing cycle [Fig.~\ref{Fig:PS}(a)].
sol:S in Fig.\ref{Fig:sol}(f) exhibits the ``swing" dynamics with $n=5$.
Compared to sol:U, $\phi\ne0$ and the initial and final states are displaced from the origin. The green vertical line in Fig.~\ref{Fig:sol}(f) indicates this.
The pendulum is flung rightwards with a finite velocity from the point displaced to the right from the bottom. It then swings back and forth before returning to the initial point [Fig.~\ref{Fig:PS}(a)].

The extremum conditions for a passive Brownian particle, which is represented by the Langevin equations Eqs.~(\ref{LE-x})--(\ref{LE-th}) with zero propulsion speed (i.e., $U=0$) are $\ddot x=\ddot y=\ddot\theta=0$, which yield only a trivial straight solution, such as sol:I, irrespective of the arbitrary boundary conditions. 
This finding emphasizes that the mobility of the ABP causes a nontrivial transition process between two states (e.g., sol:U and sol:S in Fig.~\ref{Fig:sol}).

\subsection{Periodic property of the orientation $\theta$}

\begin{figure}[t]
\begin{center}
\includegraphics[scale=0.45]{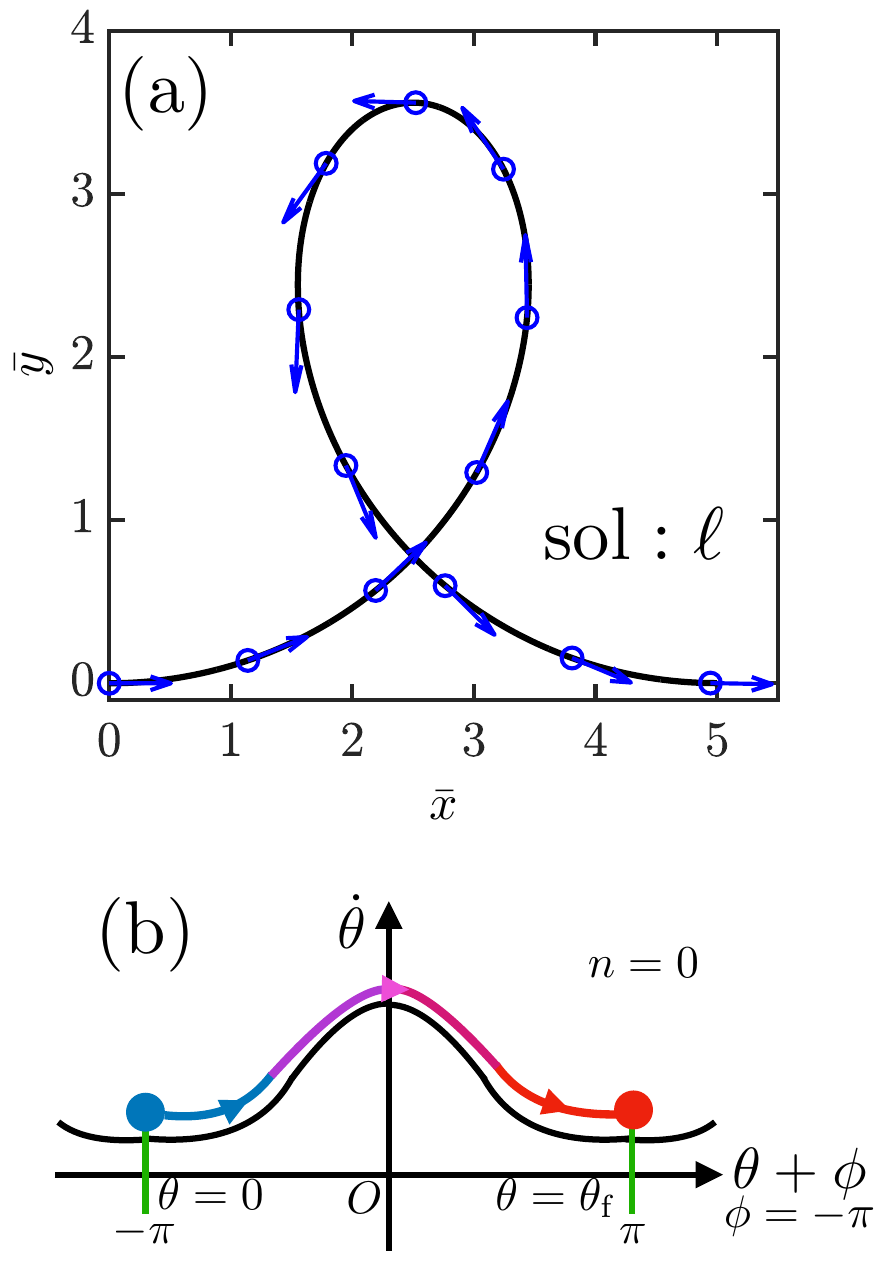}
\end{center}
\caption{Solution to Eqs.~(\ref{MPPE-x})--(\ref{MPPE-th}) with boundary conditions $\bar x_\mathrm f=5,\bar y_\mathrm f=0,\theta_\mathrm f=2\pi, t_\mathrm f/\tau=12$, which are physically similar to the final condition in Fig.\ref{Fig:sol} under the periodicity of $\theta$.
(a)	sol:$\ell$ plotted in the $\bar x$-$\bar y$ space. The blue arrows indicate $\theta$ at each point. The time interval of the arrows is $\tau$.
The parameter values are calculated as $r\approx0.1578, \phi=-\pi$ (or $\bar V_x\approx1.1578, \bar V_y=0$), and $m\approx0.8888$.
The estimated OM integral and the entropy change of the thermal bath are $\hat O=2O/(k_\mathrm BT\,\mathrm{Pe})\approx4.0534$ and $\Delta \hat s_\mathrm b=\Delta s_\mathrm b/(k_\mathrm B\,\mathrm{Pe})\approx12.4914$, respectively.
(b) Schematic showing the time evolution in the phase space of $\theta$. 
In sol:$\ell$, the behavior of $\theta$ becomes the ``rotation" dynamics, where the passage process $n=0$ only exists because $m<1$.
}
\label{Fig:soll}
\end{figure}

Due to the periodic property of the orientation, $\theta$, the final angles, $\theta_\mathrm f$ and $\theta_\mathrm f+2\omega\pi$, generate the same physical orientation, where $\omega$ is an integer (i.e., $\omega=0,\pm1,\pm2,\cdots$).
In the OM variational principle, $\omega$ is a topological rotation number indicating the number of rotations throughout the transition path from the initial to the final state.
The different rotation number, $\omega$, distinguishes the solutions of Eqs.~(\ref{MPPE-x})--(\ref{MPPE-th}) and, consequently, the locally most probable path.
Therefore, the constraint directly evaluates the OM integral to obtain the rotation number for the globally most probable path.

We now explore the most probable path of rewriting the final state as $\theta_\mathrm f=2\omega\pi$ and determine a solution with $\omega=1$, which we denote as sol:$\ell$ (Fig.~\ref{Fig:soll}(a)).
The final condition satisfied $|\theta_\mathrm f|\ge2\pi$; hence, sol:$\ell$ must be ``rotation" ($0<m<1$) with $n=0$, which is a unique solution for this condition.
Fig.~\ref{Fig:soll}(b) shows a schematic of the sol:$\ell$ dynamics in the phase space, with $\theta(t)$ monotonically increasing with time from the initial to the final state.

\subsection{Most probable path}
The solutions to the extremum conditions, Eqs.~(\ref{MPPE-x})--(\ref{MPPE-th}), are, at least, the local minimum paths. 
Therefore, by directly comparing the estimated values of the OM integral in Figs.~\ref{Fig:soll} and \ref{Fig:sol}, we deduced that the nontrivial path, sol:U, is the globally most probable path with a noticeably small OM integral, $\hat O=2O/(k_\mathrm BT\,\mathrm{Pe})\approx 3$, demonstrating the applicability of the current method with the OM integral and its variation principle.
Accordingly, sol:I, S, and $\ell$ possessed similar values of $\hat O\approx 4$, which are larger than those for sol:U.
This result may be physically interpreted by considering the relatively long final time, $t_\mathrm f/\tau=12$, to attain the position $\bar x_\mathrm f=3$. The ABP can attain the same final position in $t/\tau=3$ when there is no noise in the system; hence, it must delay by taking a detour.
We also confirmed that using simulated annealing~\cite{Kirkpatrick83,LandauBookk} for Eq.~(\ref{OMint}) makes the U-shaped path the globally most probable path.

We estimated the entropy change of the thermal bath, $\Delta s_\mathrm b$, for each solution and present its values in Figs.~\ref{Fig:sol} and \ref{Fig:soll}.
As discussed in Section\ref{OMVPsec} B, in a small noise limit, $\mathrm{Pe}\to\infty$, the averaged entropy change over the entire paths was approximated by the most probable path (i.e., sol:U) as $\langle \Delta \hat s_\mathrm b\rangle_{\mathrm i\to\mathrm f}\approx \Delta \hat s_\mathrm b^\mathrm U\approx 9$, where $\Delta \hat s_\mathrm b=\Delta s_\mathrm b /(k_\mathrm B\,\mathrm{Pe})$.
We obtained the following order of magnitude of the entropy change by comparing its values for each solution:
sol:I $<$ sol:S $<$ sol:U $<$ sol:$\ell$.
The solution with the smallest entropy change (e.g., sol:I)
does not necessarily become the most probable path.
Furthermore, we confirmed that the solution to the variational principle of the entropy change, Eq.~(\ref{EP}), only a straight path such as sol:I because $\delta (\Delta s_\mathrm b) =0$ yields $\dot{\bar x}\sin\theta_\mathrm i-\dot{\bar y}\cos\theta_\mathrm i=0$, indicating the normal velocity components' disappearance.

\section{Shape property of the most probable path for the forward translation}
\label{Sec:PhaseDiagram}
The previous section demonstrated the most probable path for translation to the front with specific parameters, namely $x_\mathrm f$ and $t_\mathrm f$.
This problem will be further discussed in this section, focusing on the shape and its dependence on parameters $x_\mathrm f$ and $t_\mathrm f$.

\begin{figure*}[!t]
\begin{center}
\includegraphics[scale=0.5]{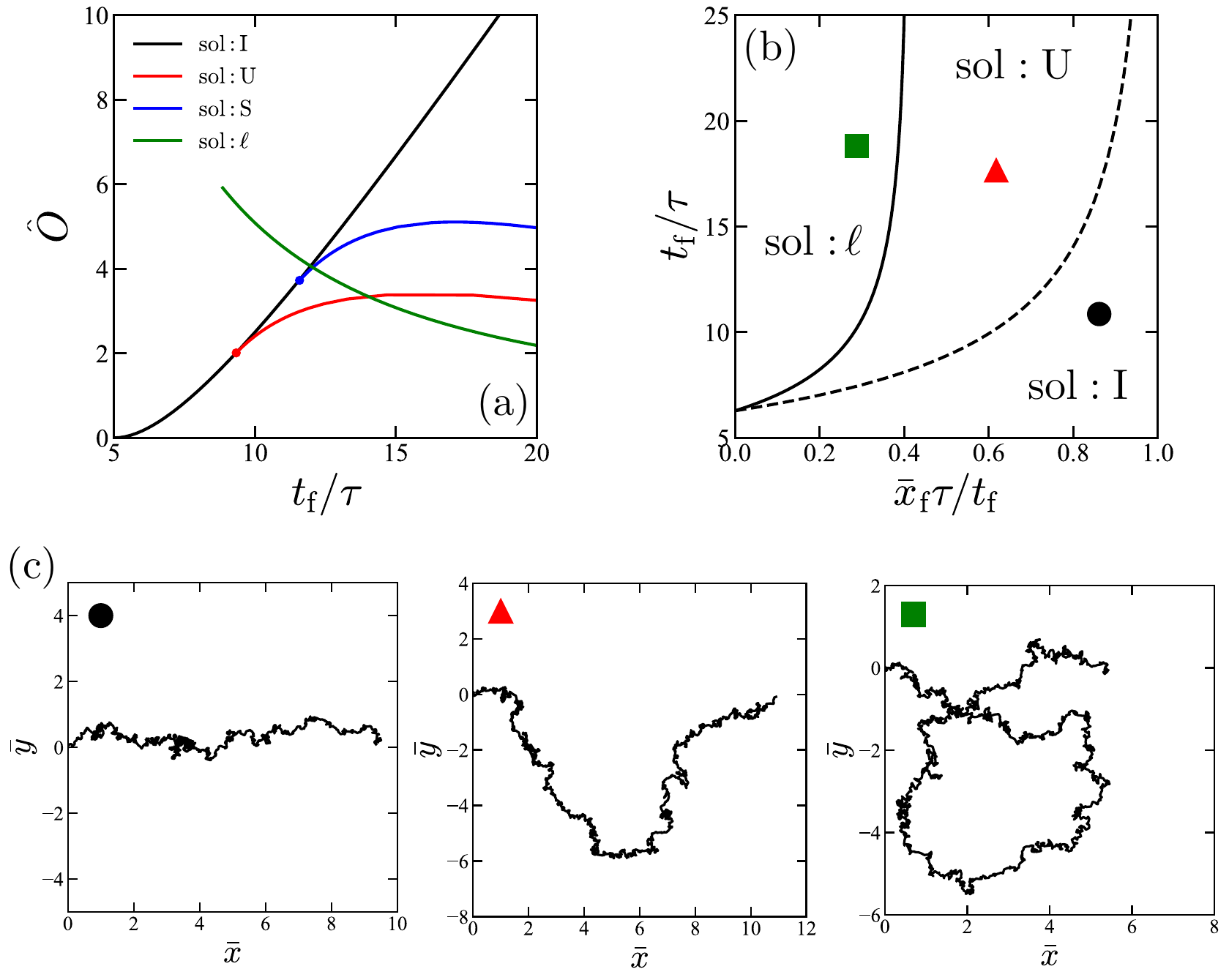}
\end{center}
\caption{
Most probable path as a function of $t_\mathrm f$ and $x_\mathrm f$ in the case of the translation to the front denoted by $\bar y_\mathrm f=0$ and $\theta_\mathrm f=2\pi\omega$, respectively, where $\omega$ is an integer. (a) Scaled OM integral, $\hat O=2O/(k_\mathrm BT\,\mathrm{Pe})$, as a function of $t_\mathrm f$ for each solution of I (black), U (red), S (blue), and $\ell$ (green) with $\bar x_\mathrm f=5$. The path with the minimum OM integral changes as I $\to$ U $\to\ell$ with an increase in $t_\mathrm f$. sol:S is always a locally minimum path lacking the smallest OM integral for the entire region. (b) Phase diagram of the globally most probable path in the parameter space spanned by $x_\mathrm f$ and $t_\mathrm f$. The solid curve indicates the boundary between sol:U and $\ell$ (i.e., $O^\mathrm U=O^\ell$). The dashed curve depicts the limiting region, where sol:U can exist, and is the boundary between sol:I and U given by Eq.~(\ref{Eq:Cond-U}). (c) Path samples satisfying $\bar y_\mathrm f=0$ and $\theta_\mathrm f=2\pi\omega$ of the Langevin simulation of Eqs.~(\ref{LE-x})--(\ref{LE-th}) with $\mathrm{Pe}^{-1}=0.08$. Each symbol (i.e., black circle, red triangle, and green square) corresponds to the parameter value shown in (b). Each random trajectory is close to the I-, U-, and $\ell$-shape.
}
\label{Fig:PC}
\end{figure*}

Fig.~\ref{Fig:PC}(a) illustrates the OM integral as a function of the final time (i.e., $O(t_\mathrm f)$) in the case of $\bar x_\mathrm f=5$ for the solutions demonstrated in Section~\ref{MPPsol}.
The black, red, blue, and green lines indicate the OM integrals for the sol:I, U, S, and $\ell$, respectively.
Fig.~\ref{Fig:PC}(a) shows that the most probable path for these solutions changed as I $\to$ U $\to\ell$ with the increasing $t_\mathrm f$. However, sol:S always possessed a larger OM integral than the other solutions for the entire region.
Furthermore, sol:U and S had lower time limits for the existence of the solution shown at the left end of the plots.
The calculation method of the OM integral for each solution is presented below and in the Appendix~\ref{App:EvaOM}.

Using the solutions to the extremum conditions shown in Eqs.~(\ref{sol-x}), (\ref{sol-y}), and (\ref{sol-th}), we recorded the OM integral, final position, and final time as $O^n(r,\phi,m)$, $x_\mathrm f^n(r,\phi,m)$, and $t_\mathrm f^n(r,\phi,m)$, respectively. Here, the possible passage process were labeled by $n$ (e.g., $n=3$ or $4$ for sol:U and $n=5$ for sol:S), as in Figs.~\ref{Fig:sol}(e) and (f).
The explicit forms of these quantities were provided in the Appendix~\ref{App:EvaOM}.
Then, we calculated the OM integral as a function of the final time, $t_\mathrm f$, under the fixed final position using these expressions for sol:I, U, S, and $\ell$. 
The value of $O(\bar x_\mathrm f=5, y_\mathrm f=0,t_\mathrm f)$ is shown in Fig.~\ref{Fig:PC}(a).

Fig.~\ref{Fig:PC}(b) depicts a sketch of the phase diagram for the most probable path shape to further clarify the shape properties in the entire parameter space.
First, we performed a simulated annealing of Eq.~(\ref{OMint}) for the entire parameter space shown in Fig.~\ref{Fig:PC}(b). Next, we obtained the I-, U-, and $\ell$-shaped paths as the globally most probable path. During this examination, the S-shaped path did not appear as the globally most probable path, which corresponded to the observation presented in Fig.~\ref{Fig:PC}(a).
The boundaries separating the parameter space for the differently shaped most probable paths are represented by the solid and dashed lines in Fig.~\ref{Fig:PC}(b) for sol:U-$\ell$ and sol:I--U and calculated as follows:

The boundary between sol:U and $\ell$ indicated that the OM integrals for sol:U and $\ell$ had the same value, i.e., $O^\mathrm U=O^\ell$.
%The OM integral values at the final position and time can be found in the Appendix~\ref{App:EvaOM}. 
We present herein the OM integral, final position, and final time as $O^\mathrm U(r,m)$, $x_\mathrm f^\mathrm U(r,m)$, and $t_\mathrm f^\mathrm U(r,m)$, respectively, for sol:U and $O^\ell(R,M)$, $x_\mathrm f^\ell(R,M)$, and $t_\mathrm f^\ell(R,M)$ for sol:$\ell$ (Appendix~\ref{App:EvaOM}).
We used $R$ and $M$ instead of $r$ and $m$ for sol:$\ell$ to distinguish it from sol:U.
Three equations, namely $O^\mathrm U(r,m)=O^\ell(R,M)$, $x_\mathrm f^\mathrm U(r,m)=x_\mathrm f^\ell(R,M)$, and $t_\mathrm f^\mathrm U(r,m)=t_\mathrm f^\ell(R,M)$, were numerically resolved for the four variables of $r$, $m$, $R$, and $M$. The solution with one degree of freedom is denoted by the solid line that separates the parameter regions of sol:U and sol:$\ell$ in Fig.~\ref{Fig:PC}(b).

We confirmed $O^\mathrm U\le O^\mathrm I$ through simulated annealing when sol:U existed for the boundary between sol:U and I.
The condition for the existence of sol:U provided the boundary of sol:U and I, which is depicted by the dashed line in Fig.~\ref{Fig:PC}(b).
We determined that sol:U only exists for the parameters satisfying the following condition:
\begin{align}
\frac{t_\mathrm f}{\tau}>\frac{2\pi}{\sqrt{1-\bar x_\mathrm f\tau/t_\mathrm f}}
\label{Eq:Cond-U}
\end{align}
This may be derived by taking the limit $m\to\infty$ for $t_\mathrm f$ and $x_\mathrm f$ of sol:U [Appendix~\ref{App:EvaOM}].
Eq.~(\ref{Eq:Cond-U}) corresponds to the dashed line in Fig.~\ref{Fig:PC} (b).

We also performed the Langevin simulation of Eqs.~(\ref{LE-x})--(\ref{LE-th}) to confirm our phase diagram and show sample trajectories, which satisfied $\bar y_\mathrm f=0$ and $\theta_\mathrm f=2\pi\omega$, in the Fig.~\ref{Fig:PC}(c). 
The parameters used in the sample paths are represented by the black circle, red triangle, and
green square symbols in Fig.~\ref{Fig:PC}(b).
The sample trajectories agreed with the most probable paths obtained using the method of the OM integral for the corresponding parameter values.

\section{Summary and Discussion}
\label{Dis}

This study analyzed the rare events of single ABP dynamics (Eqs.~(\ref{LE-x})--(\ref{LE-th})) and obtained the most probable transition process path using the OM variational principle.
First, we minimized the OM integral given by Eq.~(\ref{OMint}) to derive the extremum conditions (Eqs.~(\ref{MPPE-x})--(\ref{MPPE-th})) that must be obeyed by the most probable path.
Next, we resolved the extremum conditions with the specific initial and final conditions using analytical and numerical calculations to obtain the most probable path.

In Section~\ref{Asol}, we analyzed the extremum conditions and discovered an analogy with the pendulum motion equation in the orientation dynamics, $\theta(t)$.
The solution of Eq.~(\ref{MPPE-th}) can be classified as ``rotation" or ``swing" dynamics depending on the parameter value $m$ that indicated the inverse of the pendulum's kinematic energy.
Fig.~\ref{Fig:PS}(b) shows that these solutions can be explained using the phase space orbits spanned by $\theta$ and $\dot \theta$.
Fig.~\ref{Fig:Tp} reveals various passage processes between the two states and presents a calculation of the passage time, $T_\mathrm p^n$, for each process.
We determined the possibility that multiple solutions can be obtained from the same boundary conditions based on the $T_\mathrm p^n$ calculation.

In Section~\ref{MPPsol}, we showed the most probable path under specific, but prototypical boundary conditions (i.e., $\bar x_\mathrm f=5,\bar y_\mathrm f=0,\theta_\mathrm f=0$, and $t_\mathrm f/\tau=12$) and discovered that the system has three independent solutions: sol:I, U, and S (Fig.~\ref{Fig:sol}).
We also obtained an independent solution (i.e., sol:$\ell$) in Fig.~\ref{Fig:soll} for the condition $\theta_\mathrm f=2\pi$, which was physically similar to that of the final state, $\theta_\mathrm f=0$, because of the $\theta$ periodicity.
By estimating the OM integral for each solution directory, we conclude that sol:U is the most probable path among the four solutions of sol:I, U, S, and $\ell$.

In Section~\ref{Sec:PhaseDiagram}, we also analyzed the most probable path for the front translation with various boundary conditions satisfying $y_\mathrm f=0, \theta_\mathrm f=2\pi\omega$. 
The shape of the most probable path changed as I-, U-, and $\ell$-shape with an increasing final time, $t_\mathrm f$ (Fig.~\ref{Fig:PC}(a)).
Fig.~\ref{Fig:PC}(b) displays the trajectory shape as a phase diagram spanned by the final position and time, $x_\mathrm f$ and $t_\mathrm f$.
This phase diagram was numerically confirmed by the Langevin simulation of the original equations (Eqs.~(\ref{LE-x})--(\ref{LE-th})).

This study applied the Dirichlet (or first type) boundary conditions for the initial $x_\mathrm i,y_\mathrm i,\theta_\mathrm i$ and final $x_\mathrm f,y_\mathrm f,\theta_\mathrm f$ states.
We obtained the natural boundary conditions as follows by considering the variational principle, $\delta O=0$, at the boundaries ($t=0$ or $t=t_\mathrm f$)~\cite{CourantBook}:
\begin{align}
\dot{\bar x}\tau - \cos\theta=0,~~\dot{\bar y}\tau - \sin\theta=0,~~\dot \theta=0.
\label{BV}
\end{align}
We may use these natural boundary conditions instead of the Dirichlet boundary condition, which enable us to obtain the most probable path between two separated positions without setting the initial and final orientation, $\theta_\mathrm i$ and $\theta_\mathrm f$. 
In this case, $\theta_\mathrm i$, $\theta_\mathrm f$, and the rotation number, $\omega$, will be automatically chosen.
Even if experiments can only detect the position of an active particle rather than its orientation, the natural boundary conditions are more relevant.

Explicit calculations of the most probable path will help us understand the process in real rare transition events, such as slit passing~\cite{Salek19,Debnath21} and escape from a wall trap~\cite{Elgeti09}.
Although we focused on a free single ABP in this study, the most probable path in geometrically or mechanically confined situations (e.g., potential force~\cite{Woillez19,Gu20} and background fluid flow~\cite{Berman22}) may be calculated by employing the following equations:
\begin{align}
&\dot x=U\cos \theta-\mu\frac{\partial \Phi}{\partial x}+u_x^{\mathrm ex}(x,y)+\xi_x(t),
\label{eq-genABPx}\\
&\dot y =U\sin \theta-\mu\frac{\partial \Phi}{\partial y}+u_y^{\mathrm ex}(x,y)+\xi_y(t),
\label{eq-genABPy}\\
&\dot \theta=\Omega-\mu_\mathrm r\frac{\partial \Phi}{\partial \theta}+u_\theta^{\mathrm ex}(x,y)+\xi_\theta(t),
\label{eq-genABPth}
\end{align}
instead of Eqs.~(\ref{LE-x})--(\ref{LE-th}). Here, $\Phi(x,y,\theta)$ is the potential; $\mu$ and $\mu_r$ are the mobilities for each degree of freedom; and $u_{x,y,\theta}^{\mathrm ex}(x,y)$ is the contribution from the background fluid flow.
$\Omega$ is the chiral velocity defined as the averaged rotational velocity of particle orientation and used in chiral ABP studies~\cite{Ma22}.
We may consider an external fluid flow, $u_x^{\mathrm ex}(x,y)$, induced by the hydrodynamic interaction between particles using Fax\'en's law~\cite{Papavassiliou17, Walker22}. 
The Lagrange multiplier method can include additional constraints for the most probable path, such as spatial confinement by a wall~\cite{CourantBook}.
Another generalization is possible for individual variances,
including fluctuations in the frequencies of bacterial tumbling motions.
Therefore, individual variances may be introduced in the diffusion constant, $D_\mathrm r$.

Furthermore, the OM variational principle for multiple ABPs or continuum models of active particles can analyze rare collective events like colony splits and the dilemma of a lost child from the flock.
The OM integral for continuum fields has been proposed in the macroscopic fluctuation theory~\cite{Bertini15,Nardini17}.
This approach to rare collective events would be valuable in the field of active matter.

Another possible application of the OM integral is in optimal problems, such as travel time optimization \cite{Moreau21,Piro22}, where we selected a system's control function to minimize the target function (e.g., total travel time).
An example for this would be Eqs.~(\ref{eq-genABPx})--(\ref{eq-genABPth}) when an external potential or shear was used as the control function. The OM integral scheme can determine the optimal control, but this will be reported elsewhere.

By providing a method for determining the most probable path of an ABP with 
%a similar solution to 
the variational principle for the OM integrals, we demonstrated herein that the most probable path could be nontrivial with prototypical parameter sets using a mathematical analogy with the pendulum equation. 
%This path represents a rare transition driven by stochastic fluctuations.
Our approaches will be valuable in understanding the physical process of rare transitions, and can also be extended to more complex fluctuation-driven rare events in active matter systems.

%\begin{acknowledgment}
K.Y.\ acknowledges support by a Grant-in-Aid for JSPS Fellows (Grants No.\ 21J00096) from the JSPS. 
K.I. acknowledges the Japan Society for the Promotion of Science (JSPS), KAKENHI for Young Researchers (Grant No. 18K13456), KAKENHI for Transformative Research Areas A (Grant No. 21H05309), and the Japan Science and Technology Agency (JST), PRESTO (Grant No. JPMJPR1921).
K.Y.\ and K.I.\ are partially supported by the Research Institute for Mathematical Sciences, an International Joint Usage/Research Center located in Kyoto University.
The authors would like to thank Enago (www.enago.jp) for the English language review.
%\end{acknowledgment}

\appendix
\section{Path probability}
\label{Apps:PathProb}

\subsection{Path probability}

In this Appendix, we derive Eq.~(\ref{OMint}) by following Ref.~\cite{RiskenBook}. 
Let us consider a general system of $N$ stochastic variables, $\mathbf x(t)$, obeying the Langevin equation: 
\begin{align}
\dot x_i=v_i(\mathbf x)+\xi_i(t),
\label{App:genLE}
\end{align}
where $v_i(\mathbf x)$ ($i=1,2,\cdots,N$) is the drift velocity, and $\xi_i(t)$ is the Gaussian white noise that satisfies conditions $\langle\xi_i\rangle=0$ and $\langle\xi_i(t)\xi_j(0)\rangle =2D_{ij}\delta(t)$, where $D_{ij}$ is a diffusion tensor that is a symmetric positive definite matrix.
The Fokker--Planck equation corresponding to the Langevin equation is given as follows:
\begin{align}
\dot {\mathcal P}(\mathbf x,t)=\mathcal L(\mathbf x,t)\mathcal P(\mathbf x,t),~~\mathcal L(\mathbf x,t)=-\partial_iv_i(\mathbf x)+D_{ij}\partial_i\partial_j,
\end{align}
where ${\mathcal P}(\mathbf x,t)$ is the probability distribution functions.

The path probability, $P[\mathbf x(t)|\mathbf x^0]$, is the probability of a specific stochastic trajectory, $\mathbf x(t)$, during the time interval, $0 \le t \le t_\mathrm f$.
First, we discretized the time interval by $M$ time points as $t_0=0,t_1,\cdots,t_M=t_\mathrm f$, where the time separation is $\Delta t=t_{m+1}-t_m$.
Next, the values of the stochastic trajectory, $\mathbf x(t)$, associated with the discretized time points are given by $\mathbf x^0,\mathbf x^1,\cdots,\mathbf x^M$.
Later, we will consider the continuous representation by taking the large-$M$ limit.
The path probability, $P[\mathbf x(t)|\mathbf x^0]$, was obtained by the product of the conditional probability distribution functions, $\mathcal P(\mathbf x,t|\mathbf x',t')$, which is a solution of the Fokker--Planck equation under the initial condition ($\mathcal P(\mathbf x)=\delta(\mathbf x-\mathbf x')$ at $t=t'$) and expressed as follows:
\begin{align}
P[\mathbf x(t)|\mathbf x^0]=\lim_{M\to\infty}\prod_{m=0}^{M-1} \mathcal P(\mathbf x^{m+1},t_{m+1}|\mathbf x^{m},t_{m}).
\label{Eq:A1}
\end{align}

At a sufficiently small time separation, $\Delta t=t-t'$, the Fokker--Planck equation may be resolved as follows:
\begin{align}
\mathcal P(\mathbf x,t|\mathbf x',t')&=[1+(\Delta t)\mathcal L(\mathbf x',t')+O(\Delta t^2)]\delta(\mathbf x-\mathbf x').
\end{align}
We obtained the following using the Fourier description, $\mathbf x\mapsto \mathbf q$:
\begin{align}
\mathcal P(\mathbf x,t|\mathbf x',t')&\approx\int \frac{dq^N}{(2\pi)^N}\left[1- iq_i v_i(\mathbf x')\Delta t\right.\nonumber\\
&\left.- q_i q_j D_{ij}\Delta t\right]e^{iq_k(x_k-x'_k)}.
\end{align}
By applying the $e^{-t}\approx 1-t$ approximation, the above expression becomes
\begin{align}
\mathcal P(\mathbf x,t|\mathbf x',t')&\approx\int \frac{dq^N}{(2\pi)^N}\exp\left[- iq_i v_i(\mathbf x')\Delta t\right.\nonumber\\
&\left.- q_i q_j D_{ij}\Delta t+iq_i(x_i-x'_i)\right].
\end{align}
We obtained the following by completing the square:
\begin{align}
&\mathcal P(\mathbf x,t|\mathbf x',t')=\exp\left[-\frac{D_{ij}^{-1}}{4}V_i V_j\Delta t\right]\nonumber\\
&\times\int \frac{dq^N}{(2\pi)^N}\,e^{-D_{kl}(q_k-iD_{km}^{-1}V_m/2)(q_l-iD_{ln}^{-1}V_n/2)\Delta t},
\label{P}
\end{align}
where we introduced $V_i =(x_i-x'_i- v_i(\mathbf x')\Delta t )/(\Delta t)\approx\dot x_i- v_i(\mathbf x')$.
We only consider 
$\mathcal P(\mathbf x,t|\mathbf x',t')\sim\exp[-D_{ij}^{-1}V_i V_j\Delta t/4]$ because the Gaussian integral is a constant.

As in Eq.~(\ref{Eq:A1}), the path probability is given by the products of Eq.~(\ref{P}) and 
written as follows:
\begin{align}
&P[\mathbf x(t)|\mathbf x^0]\nonumber\\
&\sim\lim_{M\to\infty}\exp\left[-\sum_{m=0}^{M-1}\frac{D_{ij}^{-1}}{4}[\dot x_i^{m}- v_i^{m}][\dot x_j^{m}- v_j^{m}]\Delta t\right],
\label{App:PathProbDis}
\end{align}
where we define $\dot x_i^{m}=(x_i^{m+1}-x_i^{m})/(\Delta t)$ and $v_i^{m}=v_i(\mathbf x^{m})$.
We note that $v_i^{m}$ cannot be uniquely defined because of the indeterminacy of time discretization, which will be discussed in the next section.
Therefore, we obtained the following equation by introducing the integral in the exponential rather than the summation:
\begin{align}
&P[\mathbf x(t)|\mathbf x^0]\nonumber\\
&\sim \exp\left[-\int_{0}^{t_\mathrm f}dt\,\frac{D_{ij}^{-1}}{4}[\dot x_i(t)- v_i(\mathbf x(t))][\dot x_j(t)- v_j(\mathbf x(t))]\right].
\label{App:PathProb}
\end{align}
We obtained Eq.~(\ref{OMint}) with the following specific diffusion tensor and drift velocities using this expression with $N=3$:
\begin{align}
\mathbf D=
\begin{pmatrix}
D_\mathrm t &0&0 \\
0&D_\mathrm t&0\\
0&0&D_\mathrm r
\end{pmatrix},
\end{align}
\begin{align}
v_x=U\cos\theta,~~v_y=U\sin\theta,~~v_\theta=0.
\label{app:DifVel}
\end{align}

\subsection{Different definitions from the indeterminacy of time discretization}

The OM integral cannot be uniquely defined because of the indeterminacy of time discretization~\cite{Wissel79,Adib08}. 
Therefore, note that Eq.~(\ref{App:PathProb}) is only one expression of the OM integral.
We derived herein the general expression of the OM integral, including a parameter representing the time discretization method~\cite{Cates21}.
We then obtained the following equation with an expansion $v_i(\mathbf x^m+\bm \delta^m)\approx v_i^m+(\partial v_i/\partial x_j)\delta_j^m$ for small $\bm\delta^m$ and Eq.~(\ref{App:PathProbDis}):
\begin{align}
&P[\mathbf x(t)|\mathbf x^0]\sim\lim_{M\to\infty}\exp\left[-\sum_{m=0}^{M-1}\frac{D_{ij}^{-1}}{4}\right.\nonumber\\
&\times\left[\dot x_i^{m}- v_i(\mathbf x^{m}+\bm \delta^{m})+\frac{\partial v_i}{\partial x_k}\delta_k^{m}\right]\nonumber\\
&\times\left.\left[\dot x_j^{m}- v_j(\mathbf x^{m}+\bm \delta^{m})+\frac{\partial v_j}{\partial x_l}\delta_l^{m}\right]\Delta t\right]\\
&\sim\lim_{M\to\infty}\exp\left[-\sum_{m=0}^{M-1}\frac{D_{ij}^{-1}}{4}\biggl(\left[\dot x_i^{m}- v_i(\mathbf x^{m}+\bm \delta^{m})\right]\right.\nonumber\\
&\left.\left.\times\left[\dot x_j^{m}- v_j(\mathbf x^{m}+\bm \delta^{m})\right]\Delta t+2\frac{\partial v_i}{\partial x_k}\delta_k^{m}\xi_j\Delta t\right)\right]
\end{align}
where $\xi_i$ denotes the noise in the Langevin equations (Eq.~(\ref{App:genLE})).
$\bm\delta$ characterizes a time discretization method defined as $\delta_i^m=\gamma(x_i^{m+1}-x_i^m)=\gamma\dot x_i^m \Delta t$, where $\gamma$ is a parameter bounded as $0<\gamma<1$.
Hence, using the $\dot x_i\xi_j\Delta t=D_{ij}$ relation and the following previous studies~\cite{Cates21}, we obtained
\begin{align}
&P[\mathbf x(t)|\mathbf x^0]\sim\lim_{M\to\infty}\exp\left[-\sum_{m=0}^{M-1}\frac{D_{ij}^{-1}}{4}\right.\nonumber\\
&\times\biggl(\left[\dot x_i^{m}- v_i(\mathbf x^{m}+\bm \delta^{m})\right]\left[\dot x_j^{m}- v_j(\mathbf x^{m}+\bm \delta^{m})\right]\Delta t\nonumber\\
&\left.\left.+2\gamma\frac{\partial v_i}{\partial x_k}D_{k j}\Delta t\right)\right].
\end{align}
Finally, by using the small-time separation limit and replacing the summation with an integral, we obtained the general expression of the path probability as follows, including the $\gamma$ parameter:
\begin{align}
&P[\mathbf x(t)|\mathbf x^0]\sim\exp\left[-\frac{1}{4}\int_{0}^{t_\mathrm f}dt\,\biggl[D_{ij}^{-1}[\dot x_i(t) - v_i(\mathbf x(t))]\right.\nonumber\\
&\times\left.\left.[\dot x_j(t) - v_j(\mathbf x(t))]+2\gamma\frac{\partial v_i(\mathbf x)}{\partial x_i}\right]\right].
\end{align}
Notably, the calculations in the main text are independent of $\gamma$ because $\partial v_i/\partial x_i=0$ is always satisfied [Eq.~(\ref{app:DifVel})].

\section{Derivation of Eqs.~(\ref{MPPE-x})--(\ref{MPPE-th})}
\label{App:DerEq}
We derived the extremum conditions Eqs.~(\ref{MPPE-x})--(\ref{MPPE-th}) in this Appendix.
Let us consider a functional $O[\mathbf x(t)]$ of $N$ variables, $\mathbf x(t)$.
With a definition of the variations $O[\mathbf x+\epsilon \delta \mathbf x]\approx O[\mathbf x]+\epsilon \delta O+\epsilon^2\delta^2O+\cdots$, the variational principle requires the first variation of $O[\mathbf x(t)]$ to vanish (i.e., $\delta O=0$).
When we wrote the functional $O[\mathbf x(t)]$ as 
\begin{align}
O=\int_{0}^{t_\mathrm f}dt\,o(t,\mathbf x,\dot{\mathbf x}),
\end{align}
the variational equilibrium, $\delta O=0$, yielded the Euler--Lagrange equation as
\begin{align}
\frac{\partial o}{\partial x_i}-\frac{d}{dt}\left(\frac{\partial o}{\partial \dot x_i}\right)=0.
\end{align}
If a quadratic form gives the integrand $o$ as
\begin{align}
o\sim D_{ij}^{-1}[\dot x_i-v_i(\mathbf x)][\dot x_j-v_j(\mathbf x)],
\end{align}
the Euler--Lagrange equation is simplified as follows:
\begin{align}
D_{jk}^{-1}\left(\frac{\partial v_j}{\partial x_i}+\delta_{ij}\frac{d}{dt}\right)(\dot x_k-v_k)=0.
\end{align}
We obtained the extremum conditions Eqs.~(\ref{MPPE-x})--(\ref{MPPE-th}) using the specific forms of $D_{ij}$ and $v_i$ (Eq.~(\ref{app:DifVel})) for $N=3$.

\section{Extremum conditions for an ellipsoidal ABP}
\label{App:ellip}
Let us consider an ellipsoidal ABP.
The diffusion matrix considers nonsymmetric shape effects.
Therefore, using the new coordinates, $x_\parallel$ and  $y_\perp$, that move along the particle direction defined as
\begin{align}
\begin{pmatrix}
x_\parallel\\
y_\perp\\
\theta
\end{pmatrix}
=
\begin{pmatrix}
x\cos\theta+y\sin\theta \\
y\cos\theta -x\sin\theta\\
\theta
\end{pmatrix},
\end{align}
the Langevin equations of the ellipsoidal ABP are given as
\begin{align}
&\dot x_\parallel=U+\xi_\parallel(t),\\
&\dot y_\perp =\xi_\perp(t),\\
&\dot \theta=\xi_\theta(t),
\end{align}
where $U$ is the constant drift velocity and $\xi_\alpha$ is the zero-mean noise satisfying the condition $\langle\xi_\alpha(t)\xi_\beta(0)\rangle =2D_{\alpha\beta}\delta(t)$ ($\alpha=\parallel,\perp,\theta$) with the diffusion matrix:
\begin{align}
\mathbf D=
\begin{pmatrix}
D_\parallel &0&0 \\
0&D_\perp&0\\
0&0&D_\mathrm r
\end{pmatrix}.
\end{align}

We obtained the extremum conditions as follows by applying the OM variational principle (Appendix~\ref{App:DerEq}) to the original coordinates, $x,y,\theta$:
\begin{align}
&\tau\dot{\bar x}(\cos^2\theta+\lambda\sin^2\theta)+(1-\lambda)\tau\dot{\bar y}\sin\theta\cos\theta\nonumber\\
&-\cos\theta=\bar V_x-1,
\label{MPPEell-x}\\
&\tau\dot{\bar y}(\lambda\cos^2\theta+\sin^2\theta) +(1-\lambda)\tau\dot{\bar x}\sin\theta\cos\theta -\sin\theta=\lambda \bar V_y,
\label{MPPEell-y}\\
&\ddot\theta\tau=[(\lambda-1)\tau(\dot{\bar x}\cos\theta +\dot{\bar y}\sin\theta )+1][\dot{\bar x}\sin\theta-\dot{\bar y}\cos\theta],
\label{MPPEell-th}
\end{align}
which must be obeyed by the most probable paths.
Here, we introduced the length and time scales as $L=\sqrt{D_\parallel/D_\mathrm r}$ and $\tau=L/U$, respectively; further, we introduced the nondimensional positions as $\bar x=x/L$ and $\bar y=y/L$ and the aspect ratio as $\lambda=D_\parallel/D_\perp$.
$\lambda$ may vary in the range of $1/2\le\lambda\le2$ for the thermal fluctuations of a normal ellipsoidal body~\cite{KKbook}.
$\bar V_x$ and $\bar V_y$ represent the nondimensional initial velocities determined by the final conditions.
When $\lambda=1$, Eqs.~(\ref{MPPEell-x})--(\ref{MPPEell-th}) were reduced to the symmetric spherical case as  Eqs.~(\ref{MPPE-x})--(\ref{MPPE-th}).
Using Eqs.~(\ref{MPPEell-x}) and (\ref{MPPEell-y}), we rewrite Eq.~(\ref{MPPEell-th}) as follows:
\begin{align}
\ddot\theta\tau^2=-r\sin(\theta+\phi)\left[1-\frac{\lambda-1}{\lambda} r\cos(\theta+\phi)
\right],
\label{eq-pen-a}
\end{align}
where we used $r=\sqrt{(\bar V_x-1)^2+\bar V_y^2\lambda^2}$, $\cos\phi=-(\bar V_x-1)/r$, and $\sin\phi=\lambda\bar V_y/r$.
Eq.~(\ref{eq-pen-a}) is analogous to the equation of motion for a certain potential system that is not applicable for a simple pendulum and can have multiple local minima. Therefore, we obtained the following solution by multiplying Eq.~(\ref{eq-pen-a}) with $\dot\theta$ and integrating once:
\begin{align}
&\dot\theta^2\tau^2=4\frac{r}{m}\biggl[1-m\sin^2((\theta+\phi)/2)\nonumber\\
&+m\frac{\lambda-1}{\lambda} r\sin^2((\theta+\phi)/2)\cos^2((\theta+\phi)/2) \biggr],
\end{align}
where $m$ is a parameter determined by the boundary condition that can also be negative when $\lambda<1$. Using this solution, we can analyze the most probable path of the ellipsoidal particles with a method similar to that mentioned in Section~\ref{Asol}.
However, the passage process will be more complicated than the case of the simple spherical particle because the corresponding potential may have multiple local minima.

\section{Explicit form of the OM integral}
\label{App:EvaOM}
In this Appendix, we evaluated the OM integral for each solution (i.e., sol:I, U, and $\ell$) for the front translation, which required $y_\mathrm f=0$ and $\theta_\mathrm f=2\pi\omega$ ($\omega$ is an integer).
Figs.~\ref{Fig:PC}(a) and (b) were generated based on this Appendix.

\subsection{For sol:I}
Sol:I is a function of values for the boundary condition, $x_\mathrm f$ and $t_\mathrm f$, and explicitly given by $\bar x(t)=\bar x_\mathrm ft/t_\mathrm f$, $y(t)=0$, and $\theta(t)=0$.
Using Eq.~(\ref{OMint}), we obtained the OM integral for sol:I as $\hat O^{\mathrm I}=2O^{\mathrm I}/(k_\mathrm BT\,\mathrm{Pe})=(\bar x_\mathrm f\tau/t_\mathrm f-1)^2t_\mathrm f/\tau$,
which is plotted as the black curve in Fig.~\ref{Fig:PC}(a) for $\bar x_\mathrm f=5$.

\subsection{For sol:U}
Sol:U is given by Eqs.~(\ref{sol-x}), (\ref{sol-y}), and (\ref{sol-th}) with passage process $n=3$ or $4$ (Fig.~\ref{Fig:sol}(e)).
Further, we used the relation $\phi=0$, which yields $\dot{\bar x}\tau=\cos\theta-r$ and $\dot{\bar y}\tau=\sin\theta$, to satisfy $y_\mathrm f=0$. We obtained the following expression as a function of $r$ and $m>1$ by substituting these relations into the OM integral (Eq.~(\ref{OMint})):
\begin{align}
&\hat O^{\mathrm U}(r,m)=\int_0^{t_\mathrm f}\frac{dt}{\tau}\,\frac{4r}{m}\left(1-m\sin^2(\theta(t)/2)\right)+r^2 t_\mathrm f(r,m)/\tau\\
&=8\sqrt{\frac{r}{m}}\int_0^{\Theta_{\mathrm{max}}} d\theta\,\sqrt{1-m\sin^2(\theta/2)}+4r\sqrt{r} F(\pi/2,1/m),
\label{OMhatU}
\end{align}
where we changed the integrating variable $t\to\theta$ with the passage processes shown in Fig.\ref{Fig:sol}(e). $\Theta_\mathrm{max}$ is in Eq.~(\ref{Eq:thmax}). The final time is given by $t_\mathrm f^\mathrm U(r,m)=T_\mathrm c$ (Eq.~(\ref{eq-Tc})).
By defining the integrals,
\begin{align}
&G_\mathrm c(k)=\int_0^{\pi/2}dz\,\frac{\cos^2z}{\sqrt{1-k \sin^2z}},\\
&G_\mathrm s(k)=\int_0^{\pi/2}dz\,\frac{\sin^2z}{\sqrt{1-k \sin^2z}},
\end{align}
with $0<k<1$, we may simplify the expressions (\ref{OMhatU}) as
\begin{align}
\hat O^{\mathrm U}(r,m)&=16\frac{\sqrt{r}}{m}G_\mathrm c(m^{-1})+4r\sqrt{r} F(\pi/2,m^{-1}),
\label{app:OU}
\end{align}
where we used
\begin{align}
\int_0^{\Theta_\mathrm {max}}d\theta\sqrt{1-m \sin^2(\theta/2)}=\frac{2}{\sqrt{m}}G_\mathrm c(m^{-1}).
\end{align}

Introducing this into Eqs.~(\ref{sol-x}) and (\ref{sol-th}), we also demonstrated the following final position as a function of $r$ and $m$:
\begin{align}
&\bar x_\mathrm f^\mathrm U(r,m)=2\sqrt{m/r}\int_0^{\Theta_\mathrm {max}}d\theta\frac{\cos\theta}{\sqrt{1-m\sin^2(\theta/2)}}-rt_\mathrm f(r,m)/\tau\\
&=4\sqrt{r^{-1}}\left[F(\pi/2,m^{-1})-\frac{2}{m}G_\mathrm s(m^{-1})\right]-rt_\mathrm f(r,m)/\tau.
\label{app:xU}
\end{align}

The OM integral, final position, and final time were all parameterized by $r$ and $m$; thus, we calculated the OM integral for a set of $x_\mathrm f$ and $t_\mathrm f$ by tuning $r$ and $m$. 
We plotted $O^\mathrm U(\bar x_\mathrm f=5,t_\mathrm f)$ as the red curve in Fig.~\ref{Fig:PC}(a).

We considered the $m\to\infty$ limit to obtain the existing limit of sol:U described by Eq.~(\ref{Eq:Cond-U}).
As $m\to\infty$, the final position shown in Eq.~(\ref{app:xU}) became $\bar x_\mathrm f^\mathrm U=2\pi(1/\sqrt{r}-\sqrt{r})$.
Additionally, the lower time limit was obtained by taking the $m\to\infty$ limit. We then derived $t_\mathrm f^\mathrm U> 2\pi\tau/\sqrt{r}$.
Eliminating $r$ from these two equations, we obtained Eq.~(\ref{Eq:Cond-U}).

\subsection{For sol:S}
For sol:S denoted by the passage process $n=5$ shown in Fig.\ref{Fig:sol}(f), the OM integral (Eq.~(\ref{OMint})) is given as
\begin{align}
&\hat O^\mathrm S(r,m,\phi)=r^2 t_\mathrm f(r,m,\phi)/\tau\nonumber\\
&+2\int_0^{\Theta_\mathrm{max}-\phi}d\theta\,\dot\theta\tau+4\int_{-\phi}^{\Theta_\mathrm {max}-\phi}d\theta\dot\theta\tau
\label{app:OS}
\end{align}
with positive $\phi$, where we used Eqs.~(\ref{sol-x}) and (\ref{sol-y}) and changed the integrating variable $t\to\theta$ with passage processes $n=5$.

We obtained the final position as follows using the solution in Eqs.~(\ref{sol-x}) and (\ref{sol-y}):
\begin{align}
&\bar x_\mathrm f^\mathrm S(r,m,\phi)=-r\cos\phi t_\mathrm f(r,m,\phi)/\tau\nonumber\\
&+\sqrt{m/r}\int_0^{\Theta_\mathrm {max}-\phi}d\theta\,\frac{\cos\theta}{\sqrt{1-m\sin^2((\theta+\phi)/2)}}\nonumber\\
&+\sqrt{m/r}\int_{-\Theta_\mathrm {max}-\phi}^{\Theta_\mathrm {max}-\phi}d\theta\,\frac{\cos\theta}{\sqrt{1-m\sin^2((\theta+\phi)/2)}},
\label{app:xS}
\end{align}
\begin{align}
&\bar y_\mathrm f^\mathrm S(r,m,\phi)=r\sin\phi t_\mathrm f(r,m,\phi)/\tau\nonumber\\
&+\sqrt{m/r}\int_0^{\Theta_\mathrm {max}-\phi}d\theta\,\frac{\sin\theta}{\sqrt{1-m\sin^2((\theta+\phi)/2)}}\nonumber\\
&+\sqrt{m/r}\int_{-\Theta_\mathrm {max}-\phi}^{\Theta_\mathrm{max}-\phi}d\theta\,\frac{\sin\theta}{\sqrt{1-m\sin^2((\theta+\phi)/2)}}.
\label{app:yS}
\end{align}
From Eqs.~(\ref{eq-Tpf}) and (\ref{Tp4ellp1}), the final time is given as
\begin{align}
t_\mathrm f^\mathrm S(r,m,\phi)=T_\mathrm p^5=3T_\mathrm c/2-2\sqrt{m/r}F(\phi/2,m).
\label{app:tS}
\end{align}
Similar to the case of sol:U, the OM integral, final position, and final time were parameterized by $r$, $m$, and $\phi$. We then calculated the OM integral as a function of $x_\mathrm f$, $y_\mathrm f$, and $t_\mathrm f$ by tuning $r$, $m$, and $\phi$. 
Finally, we plotted $O^\mathrm S(\bar x_\mathrm f=5, y_\mathrm f=0, t_\mathrm f)$ as the blue curve in Fig.~\ref{Fig:PC}(a).

\subsection{For sol:$\ell$}
Sol:$\ell$ also obeyed Eqs.~(\ref{sol-x}), (\ref{sol-y}), and (\ref{sol-th}) with $\theta_\mathrm f=2\pi$.
In this case, we required $\phi=-\pi$, which yields relations $\dot{\bar x}\tau=\cos\theta+R$ and $\dot{\bar y}\tau=\sin\theta$ to satisfy $y_\mathrm f=0$.
In this Appendix, we used $R$ and $M$ $(0<M<1)$ instead of $r$ and $m$ to distinguish sol:U from S.
Therefore, the OM integral for sol:$\ell$ was evaluated as follows by substituting the above relations:
\begin{align}
&\hat O^\ell(R,M)=\int_0^{t_\mathrm f}\frac{dt}{\tau}\,\frac{4R}{M}\left(1-M\sin^2((\theta(t)-\pi)/2)\right)\nonumber\\
&+R^2 t_\mathrm f(R,M)/\tau\\
&=8\sqrt{\frac{R}{M}}H(M)+2R^2 \sqrt{MR^{-1}}F(\pi/2,M),
\label{app:Oell}
\end{align}
where we defined
\begin{align}
H(k)=&\int_0^{\pi/2} dz\,\sqrt{1-k\sin^2z}~~~(0<k<1),
\end{align}
and calculated the final time for sol:$\ell$ by Eq.~(\ref{eq-Tp0}) as
\begin{align}
t_\mathrm f^\ell(R,M)=T_p^0(M)=2\tau\sqrt{MR^{-1}}F(\pi/2,M).
\label{app:tell}
\end{align}
The final position was evaluated from Eq.~(\ref{sol-x}) as
\begin{align}
&\bar x_\mathrm f^\ell(R,M)=\frac{1}{2}\sqrt{M/R}\int_{-\pi}^{\pi}d\theta'\frac{\cos(\theta'+\pi)}{\sqrt{1-M\sin^2(\theta'/2)}}\nonumber\\
&+Rt_\mathrm f(R,M)/\tau\\
&=-2\sqrt{M/R}(F(\pi/2,M)-2G_\mathrm s(M))+Rt_\mathrm f(R,M)/\tau.
\label{app:xell}
\end{align}
Using a similar method in the case of sol:U, we plotted $O^\ell(\bar x_\mathrm f=5, t_\mathrm f)$ as the green line in Fig.~\ref{Fig:PC}(a) for $\bar x_\mathrm f=5$.

We generated the boundary shown as the solid line in Fig.~\ref{Fig:PC} (b) by numerically comparing $O^\mathrm U$ and $O^\ell$ in Eqs.~(\ref{app:OU}) and (\ref{app:Oell}).
On the boundary, the OM integral, final position, and final time for sol:U and $\ell$ must coincide, that is, $O^\mathrm U(r,m)=O^\ell(R,M)$, $x_\mathrm f^\mathrm U(r,m)=x_\mathrm f^\ell(R,M)$, and $t_\mathrm f^\mathrm U(r,m)=t_\mathrm f^\ell(R,M)$.
First, we numerically resolved these three equations for four variables, namely $r$, $m$, $R$, and $M$. Next, with one degree of freedom, the solution becomes the boundary in Fig.~\ref{Fig:PC}(b) separating the parameter regions of sol:U and sol:$\ell$.

%%%%%%%%%%%%%%%

\end{document}